\documentclass{aa}
\usepackage{graphicx}
\usepackage{txfonts}
\usepackage{natbib}
\bibpunct{(}{)}{;}{a}{}{,} 
\begin{document}
\title{A spectroscopic survey of the youngest field stars in the solar neighbourhood}
\subtitle{I. The optically bright sample \thanks{Based on observations collected at the \textit{Observatoire de Haute Provence} (France) and the \textit{Catania Astrophysical Observatory} (Italy)}}

\author{P. Guillout \inst{1},
	A. Klutsch \inst{1}, 
	A. Frasca\inst{2}, 
	R. Freire Ferrero\inst{1}, 
	E. Marilli\inst{2}, 
	G. Mignemi\inst{3,1}, 
	K. Biazzo\inst{2}, 
	J. Bouvier\inst{4}, 
	R. Monier\inst{5}, 
	C. Motch\inst{1},
	\and
	M. Sterzik\inst{6}  
          }

\offprints{P. Guillout\\ \email{guillout@astro.u-strasbg.fr}}

\institute{
Observatoire Astronomique, Universit\'e de Strasbourg \& CNRS, UMR 7550, 11 rue de l'Universit\'e, 67000 Strasbourg, France
\and
INAF - Osservatorio Astrofisico di Catania, via S. Sofia, 78, 95123 Catania, Italy
\and
 Dipartimento di Fisica e Astronomia, Universit\`a di Catania, via S. Sofia, 78, 95123 Catania, Italy
\and
Laboratoire d'Astrophysique de Grenoble, Universit\'e Joseph-Fourier, BP 53, 38041 Grenoble Cedex 9, France
\and
Laboratoire H. Fizeau, Parc Valrose, 28, avenue Valrose, 06108 Nice cedex 2, France
\and
European Southern Observatory, Casilla 19001, Santiago 19, Chile
}

\date{Received 7 November 2008 / accepted 27 April 2009}

 
\abstract 
{} 
{We present the first results of an ambitious ground-based observation programme conducted on 1-4 meter class telescopes. Our sample consists of  1097 active and presumably young stars, all of them being optical (Tycho Catalogue) counterparts  of ROSAT All-Sky Survey X-ray sources in the northern hemisphere. In this paper, we concentrate on the optically brightest ($V_{\rm T}\leqslant$~9$\fm$5) candidates (704 objects). We acquired high-resolution optical spectroscopy in the H$\alpha$ and/or lithium spectral regions for 426 of such stars without relevant data in the literature. We describe the star sample and the observations and we start to discuss the physical properties of the investigated stars.}
 {We used a cross-correlation technique and other tools developed by us to derive accurate radial and rotational velocities and to perform an automatic spectral classification for both single stars and double-lined systems. The spectral subtraction technique was used to derive chromospheric activity levels and lithium abundances. We estimated the fraction of young single stars and multiple systems in stellar soft X-ray surveys and the  contamination by more evolved systems, like RS~CVn binaries. We classified stars on the basis of their lithium abundance and give a glimpse of their sky distribution.}
 {The sample appears to be a mixture of quite young \textit{Pleiades-like} and \textit{Hyades-like} stars plus an older lithium-poor population probably born within the last 1-2 Gyr. Seven stars with a lithium abundance compatible with the age of IC~2602 (about 30~Myr) or even younger were detected as well, although two appear to be lithium-rich giants.The discovery of a large number of highly or moderately lithium-rich giants is another outcome of the present survey.}
 {The contamination of soft X-ray surveys by old systems in which the activity level is enhanced by tidal synchronisation is not negligible, especially for K-type stars. Five stars with lithium content close to the primordial abundance are probably associated with already known moving groups in the solar neighbourhood. Some of them are good post-T Tauri candidates according to their positions in the HR diagram.}

   \titlerunning{A Spectroscopic Survey of Youngest Field Stars in the Solar Neighbourhood}
      \authorrunning{P. Guillout et al.}

\maketitle

\section{Introduction}
\label{Sec:intro}

Even if the global scenario of star formation history of the Milky Way is fairly well known from its birth until the last few giga-years \citep{1997A&A...320..440H}, surprisingly the recent local star formation rate is hardly constrained. Although newborn stars are usually confined to the locations of their parental clouds and share the same region of space (i.e. open clusters or associations), several processes tend to disperse them. Ejection of unused gas can reduce the mass of a cluster so much that it becomes gravitationally unbounded and breaks up. Moreover, occasional close encounters with giant molecular clouds along galactic rotation may even rapidly disrupt the cluster entirely. As a result, most open clusters only survive a few hundred million years, although the cluster lifetime sensitively depends on the cluster's mass (for example M67 is a cluster with a solar age). Once mixed in the galactic plane environment, young stars are virtually indiscernible because neither their global photometric properties nor the presence of nearby gas can help to disentangle them from older stars. As a consequence, the recent (i.e. last giga-years) local star formation rate is poorly known because of the difficulties encountered in properly selecting young main sequence (MS) late-type stars in the field from optical data alone. The same problem holds for the scale height of stars.  As stars get older they tend to scatter higher above the galactic plane, a process known as disc heating. The thermalisation of the young stellar population is still an open question, and their scale height is generally assumed to be similar to the one of the molecular clouds from which they are born. 

Fortunately, the situation has improved by making use of X-ray observations. The discovery that stars of almost all spectral types were X-ray emitters with luminosities in the range 10$^{26}$\,-\,10$^{34}$\,erg s$^{-1}$ was one of the major results of the Einstein mission \citep{1981ApJ...245..163V}. The current knowledge of the composition of the soft X-ray sky is mainly based either on complete optical identifications of ROSAT All-Sky Survey (RASS; \citealt{1999yCat.9010....0V, 2000IAUC.7432....3V}) sources in small-size regions (\citealt{1997A&A...318..111M, 1997A&AS..123..103Z, 2005A&A...433..151Z}) or on the cross-correlation of the RASS with Tycho and Hipparcos catalogues \citep{ESA1997}. These last ones (namely RasTyc and RasHip; \citealt{1999A&A...351.1003G}) allowed study of the large-scale (all-sky) distribution of X-ray active stars in the solar neighbourhood \citep{1998A&A...334..540G} and led to the discovery of the late-type stellar population of the Gould Belt (GB; \citealt{1998A&A...337..113G}).

As chromospheric and coronal activity levels decrease with increasing age, they can serve as proxy age indicators (\citealt{1996AJ....111..439H, 1997AJ....114.1673S}). During the last decade a fascinating picture of the recent star formation history in the solar neighbourhood has emerged. About fifty Myr ago, molecular clouds were forming stars near the present position of the Sun. Thirty Myr ago, the first generation of massive stars exploded as supernovae triggering star formation in an expanding ring-like structure and modelled the gas cavity in the solar neighbourhood (i.e. the local Bubble). Although the present rim coincides with most of the nearby OB associations, the distribution of young stars as outlined by the RasTyc sample indicates that stellar formation also took place also inward the Belt, over a significant, yet poorly constrained, radial extent (the lack of active star formation within 50~pc of the Sun is believed to be due to the local Bubble). Although the massive-star content has been extensively studied, less is known about low-mass star formation in the Belt. Its age is uncertain by a factor of 2 because of the discrepancy found between the dynamical time scale (20-30~Myr; \citealt{1999ApJ...522..276M, 2003A&A...404..519P}) and the stellar age (30-60~Myr; \citealt{1994A&A...281...35C, 2000A&A...359...82T}).

It is known from statistical studies that open clusters account for only about 10\% of the total galactic star formation.
However, \citet{Zinnecker2008} and  \citet{2008A&A...487..557P} have recently suggested that it may account for up to 50\%, the remainder occurring in OB associations. Nevertheless, isolated very young stars have recently been detected. Although TW Hya displays all the characteristics of T Tauri stars (TTS), it has earned notoriety mostly on the basis of the absence of dark clouds in its vicinity. Different scenarios have been proposed as possible explanations for the presence of TTS far away from molecular clouds. \citet{1995A&A...304L...9S} suggested that isolated pre main-sequence (PMS) stars have been ejected ({\it run-away} TTS) from their birthplace as a consequence of close encounters with other members of their parent cloud. Alternatively, they may have been formed locally from small clouds, as suggested by \citet{1996ApJ...468..306F}. As an isolated TTS, the origin of TW Hydra has long been a mystery for astronomers. However, searches by \citet{1989ApJ...343L..61D} and \citet{1992AJ....103..549G} have revealed four other TTS in the same region of the sky that shares the same space motion. At a distance of only $\thickapprox$~50~pc and an age $\thickapprox$~12~Myr, TW Hya is now recognised as one of the closest known regions of recent star formation, i.e. the TW Hya association \citep{2000ApJ...535..959Z}. Since its discovery, eight more young nearby associations have been identified so far \citep{2006ApJ...649L.115Z}, and even more  are suspected on the basis of kinematics or other features \citep{2008arXiv0808.3362T}.\\

Another issue regarding young stars is known as the post-T Tauri star (PTTS) problem. PTTSs cover the evolutionary period between 10$^{6}$ to 10$^{7}$ years (i.e. between the T Tauri phase and zero-age main sequence ZAMS age). They are expected to be abundant in the vicinity of star formation regions, although studies that concentrate on such regions often fail to find the number of PTTS that are expected from star formation and evolution theories, as already noted by \citet{1978ppeu.book..171H} thirty years ago. \citet{1998ApJ...498..385S} discussed possible explanations of the PTTS problem and in particular suggested that numerous PTTS may exist in isolated environments, but they are difficult to recognise because they mostly lack the obvious optical signatures of their T Tauri progenitors.\\

Given their age and proximity, young, nearby associations and PTTS are important for understanding the evolution of circumstellar discs (in particular the transition phase of disc-dispersal) and the formation of planets, as well as for giving new insight into the process of planet migrations. RasTyc and RasHip samples may contain hundreds of such hidden young stars, and complementary optical data are absolutely necessary if the huge scientific potential of these samples is to be fully exploited. Moreover, detailed comparisons of radial velocity, proper motions, age, spectral type, and luminosity class distributions with predictions of stellar X-ray population model \citep{1996A&A...316...89G} are likely to establish new constraints on galactic evolution parameters such as the recent local stellar formation rate and the scale height of young stars.\\

With this scientific background in mind, we started an optical high-resolution spectroscopic campaign of a large sample of RasTyc~/~RasHip stars on 1- to 4-meter class telescopes. The sample is mostly composed of G and K late-type stars with Tycho $B-V$ colour index in the range 0.6-1.33. The present paper is dedicated to the optically bright sample (composed of stars with $V_{\rm T}$~$\leqslant$~9$\fm$5) for which we already have a complete dataset. We describe the observations and discuss stellar properties as derived from high-resolution spectra. An outstanding result is the discovery of new members of moving groups and possible PTTS candidates, as well as the detection of several lithium-rich giants. We note that a complementary study in the southern hemisphere has been conducted by \citet{2006A&A...460..695T}. In the following we first introduce our stellar sample, the observations, and the reduction methods (Sect.~\ref{Sec:Data}), then we present our determination of the stellar parameters (Sect.~\ref{Sec:APs}), the level of chromospheric activity, and the age evaluation (Sect.~\ref{Sec:ActAg}). In Section~\ref{Sec:Disc} we discuss the fraction of multiple systems, evolutionary status, age, and moving group membership. We conclude and summarise the first results from the ongoing study of these data (Sect.~\ref{Sec:C&P}).
%
\section{Sample selection and observing campaign}
\label{Sec:Data}

As a starting point we used the RasTyc sample resulting from the cross-correlation of the RASS with the Tycho catalogue. To obtain statistically significant results and to fulfil our scientific goals, we selected about 1100 RasTyc sources in the solar neighbourhood given the following constraints:
\begin{itemize}
\item declination $\delta \geqslant 0^{\circ}$ ensuring that any target can be observed at low airmass from French and Italian national telescopes;
\item right ascension $\alpha \in [15~H, 8~H]$ to cover both low and high galactic  plane regions observable from the May-June to December-January periods decided as our observing slots in the framework of the \textit{Observatoire de Haute Provence} (OHP - France) key programme;
\item $ 0.6~\leqslant~(B -V)_{\rm T}~\leqslant~1.3$ range for which the lithium depletion behaviour is more effective. Indeed, it is well known that  the strength of the \ion{Li}{i}\,$\lambda$6707.8 line can be used to indicate youth \citep{1996A&A...306..408M} for stars cooler than about  mid-G, although for M-type stars, the lithium is burned so rapidly in their deep convective envelopes that the \ion{Li}{i}\,$\lambda$6707.8  line is only detectable for extremely young stars. The lithium plateau observed for F-type stars does not permit the use of this line as an age diagnostic;
\item ROSAT PSPC count rate greater than or equal to 0.03~cts\,s$^{-1}$ to avoid RASS scanning bias (see \citealt{1999A&A...351.1003G});
\item V$_{\rm T}~\leqslant$~10$\fm$5, since Tycho is largely incomplete above this limit.
\end{itemize}
\begin{figure}[t]
\includegraphics[width=7.0cm,bb=50 30 470 664,angle=90]{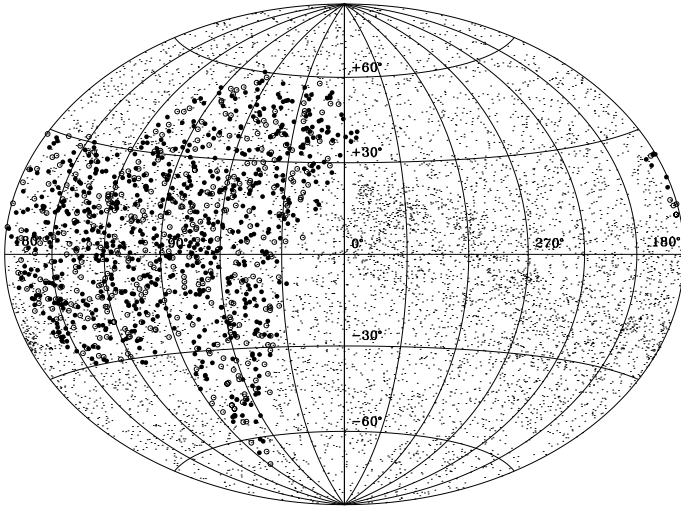}
\caption{Galactic distribution (Aitoff projection) of the stars selected for this spectroscopic survey (big circles) overlaid on the 8593 RasTyc stars with PSPC count-rate larger or equal to 0.03~cts~s$^{-1}$ (small dots). Only the optically bright sample (filled circles) is investigated in the present paper. The remainder (faint sample; open circles) will be studied in a subsequent paper.}
\label{Fig:SkyDistr}
\end{figure}

\begin{figure*}[t]
\includegraphics[width=9.0cm]{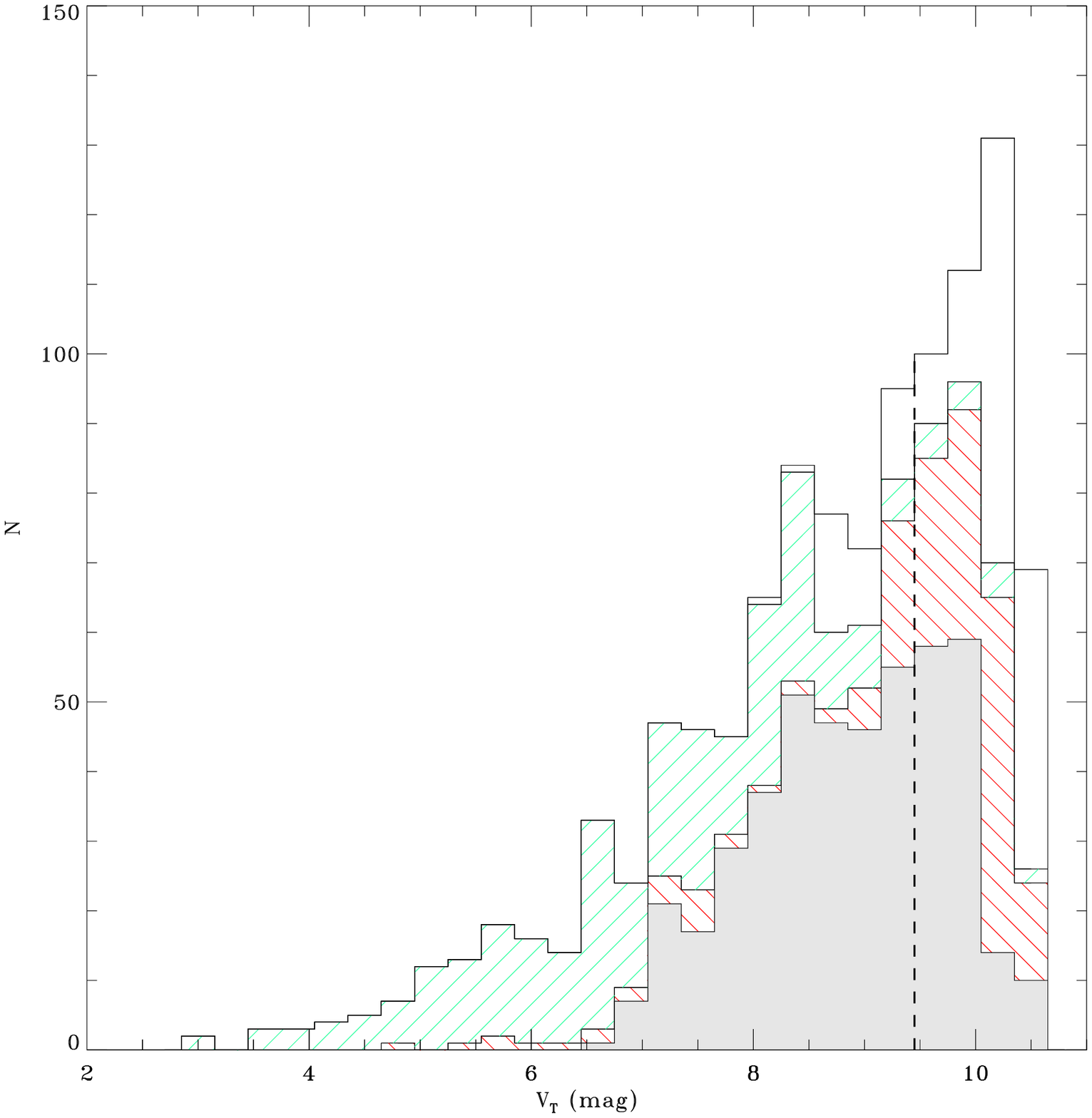}
\includegraphics[width=9.0cm]{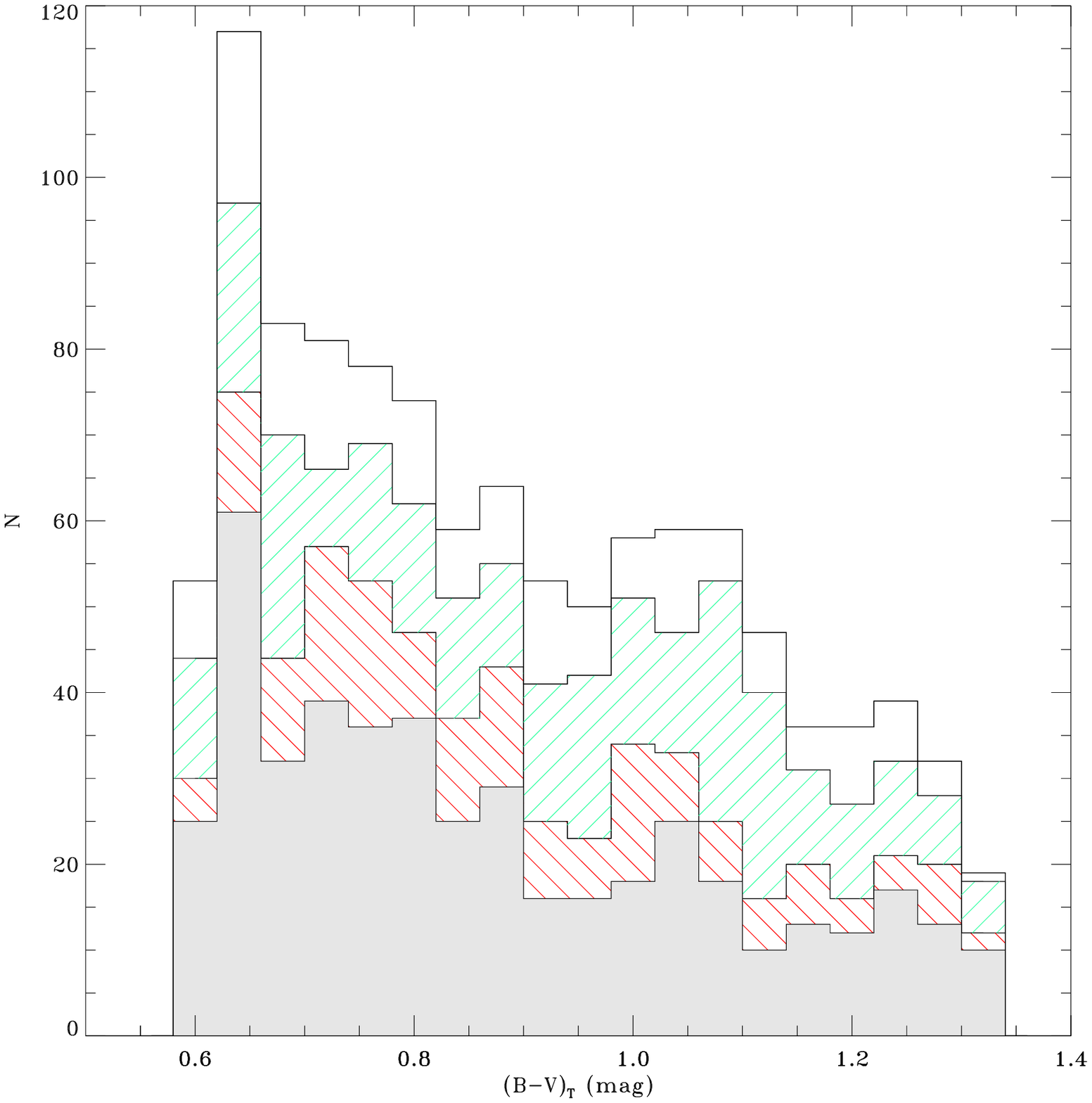}
\caption{Tycho $V_{\rm T}$ magnitude {\it (left panel)}  and $(B-V)_{\rm T}$ colour {\it (right panel)} distributions of the stars in our sample. Stars with already available relevant measurements from the literature (green right-stripped area) have been systematically discarded from the observation programme. Spectroscopic observations of the optically \textit{bright sample} have been conducted mostly on the OHP 1.52-m telescope (grey area), while a significant part of the optically \textit{faint sample} has been scheduled on the larger OHP 1.93-m and TNG 3.58-m telescopes (red left-stripped area). A small number of stars (white area ; $V_{\rm T}$~$\leqslant$~9$\fm$5) have also been scheduled on the OAC 0.91-m telescope.}
\label{Fig:MagCol}
\end{figure*}

We show in Fig.~\ref{Fig:SkyDistr} the galactic distribution of the make-up sample overlaid on the RasTyc all-sky picture, while Fig.~\ref{Fig:MagCol} displays the selected stars magnitude and colour distributions. The median visual magnitude of the sample is $\approx$~9$\fm$0. To gather as much as possible a-priori information on these potential targets and to minimise observing time, the data from the literature were checked against the SIMBAD database. Stars brighter than $V_{\rm T}\thickapprox$~6$\fm$0 mostly have over 50 bibliographic references with at least one spectroscopic investigation so were not scheduled for the observations. Those fainter than this limit were checked one by one and discarded only if relevant spectroscopic data already existed. Nevertheless, a few stars in common with the KPNO survey \citep{2000A&AS..142..275S} have been observed to test the consistency of our results with previous works.\\

The main observation programme was conducted mostly in service mode at the OHP 1.52-m telescope from November 2001 to August 2005. We used the \textit{Aurelie} spectrograph \citep{1994A&AS..108..181G} in both the H$\alpha$ and lithium spectral regions with grating 7 (18~000 grooves/mm), which allowed us to cover over 120~$\AA$ at a resolution of $R\approx$~38\,000. To derive reliable equivalent widths and good radial and rotational velocities, a signal-to-noise ratio (S/N) of 70-100 per resolution element (depending on the star magnitude) was required. The \textit{Horizon 2000} CCD in bin mode permitted us to reach our requirements in the most efficient way.\\ 
Another set of 110 stars observed in 2000 and 2001 (for a similar research project but over a smaller area) at the OHP 1.93-m telescope was merged with the previous sample. Spectra covering the 3900-6800 $\AA$~range were acquired with the \textit{Elodie} \'echelle spectrograph \citep{1996A&AS..119..373B}  with a resolving power of $R\approx$~42\,000. The equipment of both telescopes has about the same efficiency. The typical exposure times ranged from less than 10~minutes for a bright 7$\fm$0 star up to an hour for stars fainter than 9$\fm$0, for both instruments.
Finally, 45 remaining stars brighter than $V_{\rm T}$~=~9$\fm$0, which could not be scheduled at the OHP because of  time constraints or bad weather conditions, have been successfully observed at the $M. G. Fracastoro$ station (Serra  La Nave, Mt. Etna, 1750 m a.s.l.) of the \textit{Osservatorio Astrofisico di Catania} (OAC - Italy). The OAC 0.91-m telescope was equipped with the FRESCO fiber-fed \'echelle spectrograph covering the spectral range 4250-6850\,\AA\  with a resolution of $R\simeq\,21\,000$. Spectra of radial and rotational velocity standard stars, as well as bias, flat-field, and arc-lamp exposures, were acquired at least twice a night for calibration purposes. In total $\approx$~1400 spectra were acquired during the course of the $\approx$~190 nights devoted to the programme (Table~\ref{Tab:Obs_Sum}) with various instrumental setups (Table~\ref{Tab:Inst_Sum}).
\begin{table}[t]
\caption{Observation program summary}
\begin{tabular}{ccccc}
\hline \hline
Year  & Period & Nights & Telescope & Spectra \\ 
\hline
2000 & July~-~September & 22 & OHP 1.93-m & 60 \\
2001 & August & 12 & OHP 1.93-m & 50 \\
2001 & October~-~December & 29 & OHP 1.52-m & 290 \\ 
2002 & June~-~November  & 53 & OHP 1.52-m & 396 \\
2003 & January & 23 & OHP 1.52-m & 207 \\ 
2004 & June~-~September & 29 & OHP 1.52-m & 231 \\ 
2005 & July~-~August & 20 & OHP 1.52-m & 126 \\
2007 & September~-~October & $^{*}$ & OAC 0.91-m & 28 \\ 
2008 & March~-~July & $^{*}$ & OAC 0.91-m & 18 \\ 
\hline
Total &  & 188+ &  & 1~406 \\ 
\hline 
\end{tabular}
\begin{list}{}{}
\item[$ ^{*}$] Queue mode.
\end{list}
\label{Tab:Obs_Sum}
\end{table}
\begin{table}
\caption{Instrumental setup summary}
\begin{tabular}{lllcc}
\hline \hline
Telescope  & Spectrograph & Spectra & Spectral region &Resolution \\ 
\hline
OHP 1.93-m & \textit{Elodie} & \'echelle & 3900-6800 $\AA$ & 42\,000 \\
OHP 1.52-m & \textit{Aurelie} & 1D & 6495-6625 $\AA$ & 38\,000 \\ 
OHP 1.52-m & \textit{Aurelie} & 1D  & 6650-6775 $\AA$ & 38\,000 \\ 
OAC 0.91-m & FRESCO & \'echelle & 4250-6850 $\AA$ & 21\,000 \\ 
\hline 
\end{tabular}
\label{Tab:Inst_Sum}
\end{table}

The 1D \textit{Aurelie} spectra were corrected for bias, flat-field, and cosmic rays hits by using standard MIDAS procedures while, the OAC data reduction was performed by using the ECHELLE task of IRAF\footnote{IRAF is distributed by the National Optical Astronomy Observatories, which are operated by the Association for research in Astronomy, Inc. under cooperative agreement with  the National Science Foundation.}. \textit{Elodie} spectra were automatically reduced online during the observations with the  standard pipeline reduction package, which also includes the cross-correlation with a reference mask and the TGMET facility for spectral classification \citep{1998A&AS..133..221S}. Telluric lines were subtracted following the procedure described by  \citet{2000A&A...364..179F}, and spectra were then normalised to the continuum through a polynomial fit avoiding  spectral regions heavily  affected by blends of absorption lines, molecular bands, and the  H$\alpha$ wings. Spectra were finally corrected for the Earth velocity and reduced to the heliocentric rest frame.

\section{Determination of astrophysical parameters}
\label{Sec:APs}

\subsection{Multiplicity, radial and projected rotational velocity}
\label{Sec:MRvVsini}
\begin{figure*}[t]
\includegraphics[width=6.0cm]{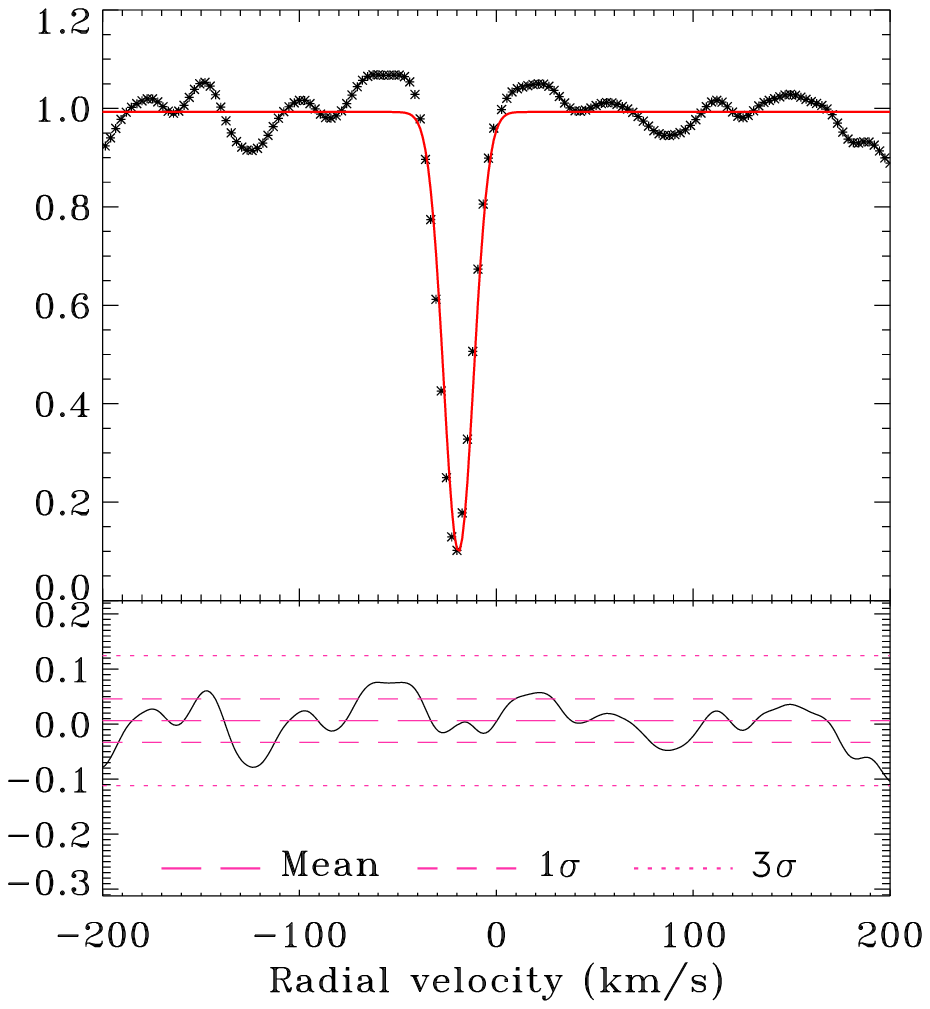}
\includegraphics[width=6.0cm]{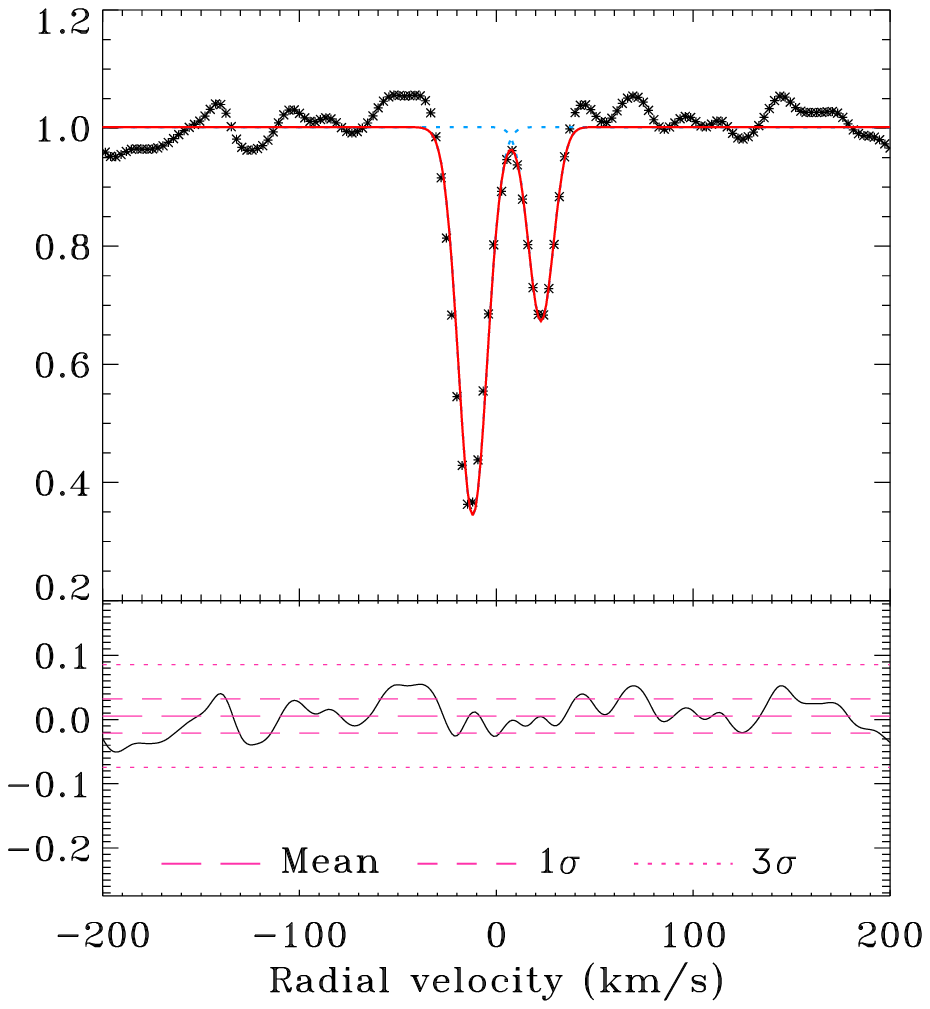}
\includegraphics[width=6.0cm]{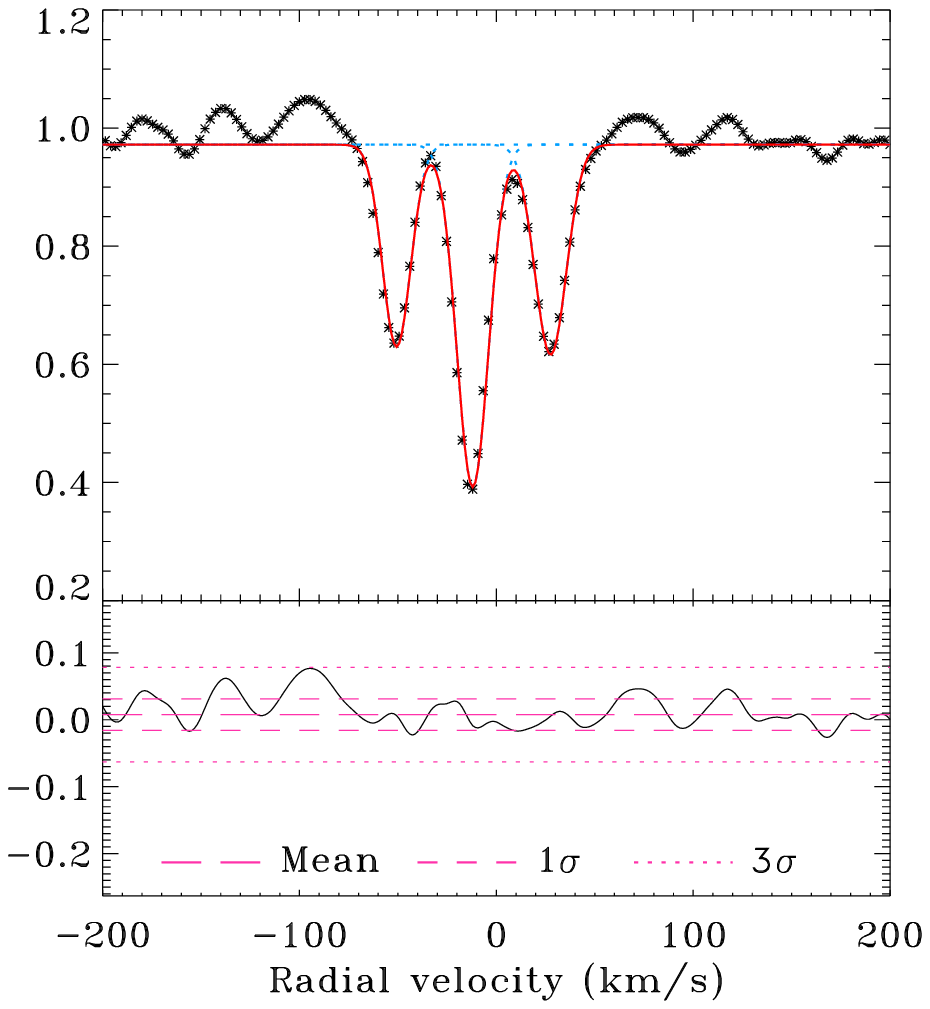}
\caption{Cross-correlation function (asterisks) and Gaussian fitting (red solid line) for the single star HD 6026 (RasTyc~0102+6236; \textit{left panel}), the SB2 binary BD+67 34 (RasTyc~0025+6848; \textit{middle panel}), and the triple  system BD+82 622 (RasTyc~2034+8253; \textit{right panel}). Lower panels show residuals.}
\label{Fig:CCF}
\end{figure*}

For each star observed with \textit{Aurelie}, we computed the cross-correlation function (CCF) independently for both the H$\alpha$ and lithium spectral regions by cross-correlating the observed spectra with a synthetic mask. Since the CCF improves when spectra with a very similar shape are correlated, we used a grid of 9 different synthetic spectra (corresponding to F0, F5, G0, G5, K0, K5, M0, and M5 spectral types) as CCF masks, and selected the one that is more similar to the spectral type (or the $B-V$ colour index) of the target. The CCF is a powerful tool for revealing single, double, and multiple systems. Significant minima in the CCF were searched for automatically and single, double, or multiple functions (i.e. up to 4 Gaussian or rotational profiles) were fitted to the CCF dips (see Fig.~\ref{Fig:CCF}). The standard deviation of the CCF values far from the central dip(s) ($\sigma$) was evaluated for each spectrum and all the minima deeper than 3$\sigma$ were kept as significant. This allowed us to sort out single stars (S), double-lined binaries (SB2), triple systems (SB3), and multiple stars (M).\\
\begin{figure}[t]
\includegraphics[width=9.0cm]{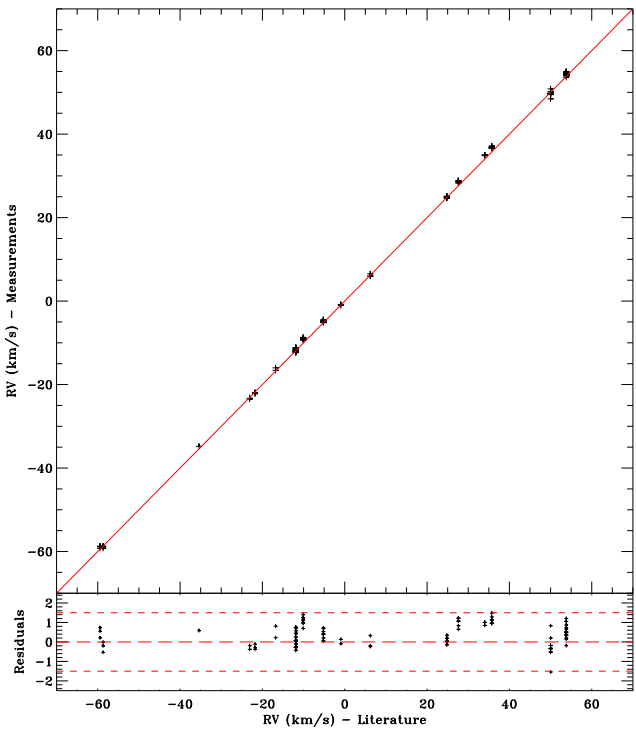}
\caption{Comparison of our radial velocities measurements with those quoted in the literature for the standard stars observed during our runs. Residuals from the one-to-one relation are shown on the bottom panel. The short-dashed lines delimit a $\pm$~1.5~km~s$^{-1}$ region around the one-to-one relation (continuous line in the upper box).}
\label{Fig:RVCalib}
\end{figure}

In principle, the minimum of the CCF readily gives the radial velocity (RV) of the star. However, the CCF profile may appear asymmetric or distorted because of cool star-spots on the stellar photosphere. Therefore, we fitted the entire CCF profile to obtain a more accurate RV determination. 

For most of our spectra, the RV uncertainty is dominated by the accuracy of the wavelength calibration. The effects of spectrograph drift during the night were minimised by acquiring arc lamp exposures at least twice per night. An overall RV precision of $\pm$~1.5~km\,s$^{-1}$ was estimated by comparing our RV measurements with those from the literature for a set of standard stars (see Fig.~\ref{Fig:RVCalib}) although the precision can reach 0.5~km\,s$^{-1}$ for high S/N spectra.\\ 

\begin{figure}[t]
\includegraphics[width=9.0cm]{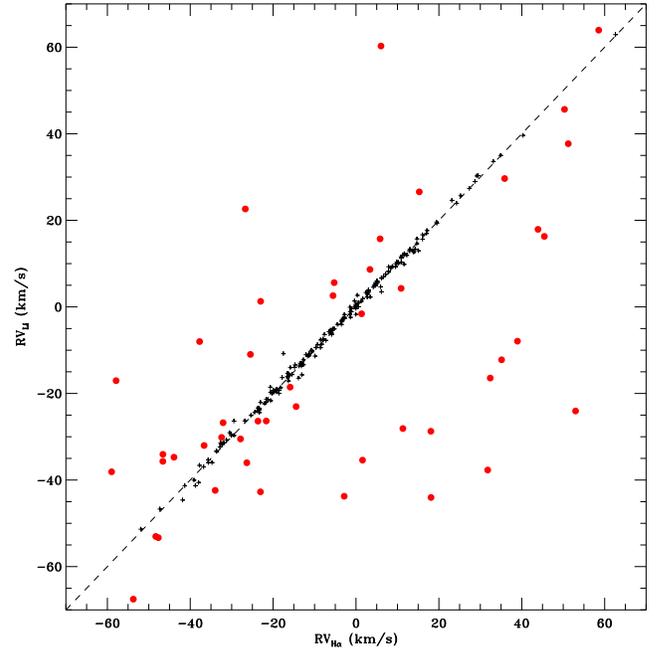}
\caption{Radial velocities measured from the lithium (RV$_{Li}$) versus H$\alpha$ (RV$_{H_{\alpha}}$) spectral regions for stars classified as single from the CCF profile analysis. The dashed line shows the one-to-one relation. Stars with inconsistent RV$_{Li}$-RV$_{H_{\alpha}}$ measurements (filled red circles) are interpreted as single-lined spectral binaries (SB1).}
\label{Fig:SB1}
\end{figure}

We also took advantage of each star normally being observed a few nights apart, in the H$\alpha$ and \ion{Li}{i} regions, to gather spectroscopically unresolved  binaries. For stars appearing as singles, i.e. with single-lined spectra without any detectable RV variation on a timescale of a few days, we averaged the two RV values. On the other hand, significant RV differences between the H$\alpha$ and lithium spectra were interpreted as real variations. Thus, the stars with a single dip in the CCF and a variable RV were classified as single-lined binaries (SB1; see Fig.~\ref{Fig:SB1}) whenever the differences in RV measurements were greater than the sum of their uncertainties.\\

The CCF also permits the projected rotational velocity ($v\sin i$) to be determined. The shape of the CCF is sensitive to the rotation speed. At low $v\sin i$, i.e. when the spectral lines are mainly affected by the Doppler and microturbulence broadenings, a Gaussian function generally gives a very good approximation of the CCF profile and we used a calibration relation between the width of the fitted Gaussian, $\sigma$(CCF), and $v\sin i$ to determine the rotational velocity, following a method similar to that of \citet{1998A&A...335..183Q}. At high $v\sin i$ values, the shape of the CCF is far from Gaussian and becomes more and more like a rotational profile. In such cases, a Gaussian function would overestimate the rotational velocity, so we preferred to compare the observed spectra with a grid of synthetic templates broadened by the convolution with a rotation profile and to retain the $v\sin i$ giving the lowest residuals in the difference.

The error on the projected rotational velocity was computed for each star and depends critically on S/N ratio, as well as on the shape of the $\sigma$(CCF) versus $v\sin i$ relation. At low $v\sin i$ ($\lesssim$~25\,km~s$^{-1}$), the relation can be approximated with a square-root function and a small error on the $\sigma$(CCF) propagate on a significant $v\sin i$ relative error (50~\%). On the other hand, at high $v\sin i$, the relation is rather linear and the relative error is lower (10~\%). In any case, our spectral resolution does not permit to derive $v\sin i$ values lower than about 5~km\,s$^{-1}$. Finally, one must keep in mind that  unresolved lines from a secondary companion can mimic a higher projected rotational velocity (and can bias the estimation of the astrophysical parameters).\\

For the 45 stars observed with FRESCO at OAC, we adopted spectra of F-G-K radial velocity standard stars acquired during the observing runs as RV templates for the cross-correlation. We used all the 20 \'echelle orders, excluding the H$\alpha$ and \ion{Na}{i} D$_2$ lines from the analysis, because of their strong wings affecting the CCF shape and due to the contamination of these lines by chromospheric emission. The radial velocity measurements, performed with the IRAF task {\sc Fxcor}, were obtained by averaging RV data from all \'echelle orders with the usual instrumental weight $w_{\rm i}=\sigma_{\rm i}^{-2}$. The $\sigma_{\rm i}$ values were computed by {\sc Fxcor} according to the fitted peak height and the antisymmetric noise as described by \citet{1979AJ.....84.1511T}. The standard errors of the weighted means were computed on the basis of the errors $\sigma_{\rm i}$ in the RV values for each order according to the usual formula \citep[see, e.g.,][]{1957AmJPh..25..498T}. For the $v\sin i$ determination, we used the same method as for the \textit{Aurelie} spectra but the spectra of slowly-rotating standard stars were used as templates.

\subsection{Effective temperature, gravity, and metallicity}
\label{Sec:TeffloggFeH}
\begin{figure}[t]
\includegraphics[width=9.0cm,bb=25 10 453 260,clip]{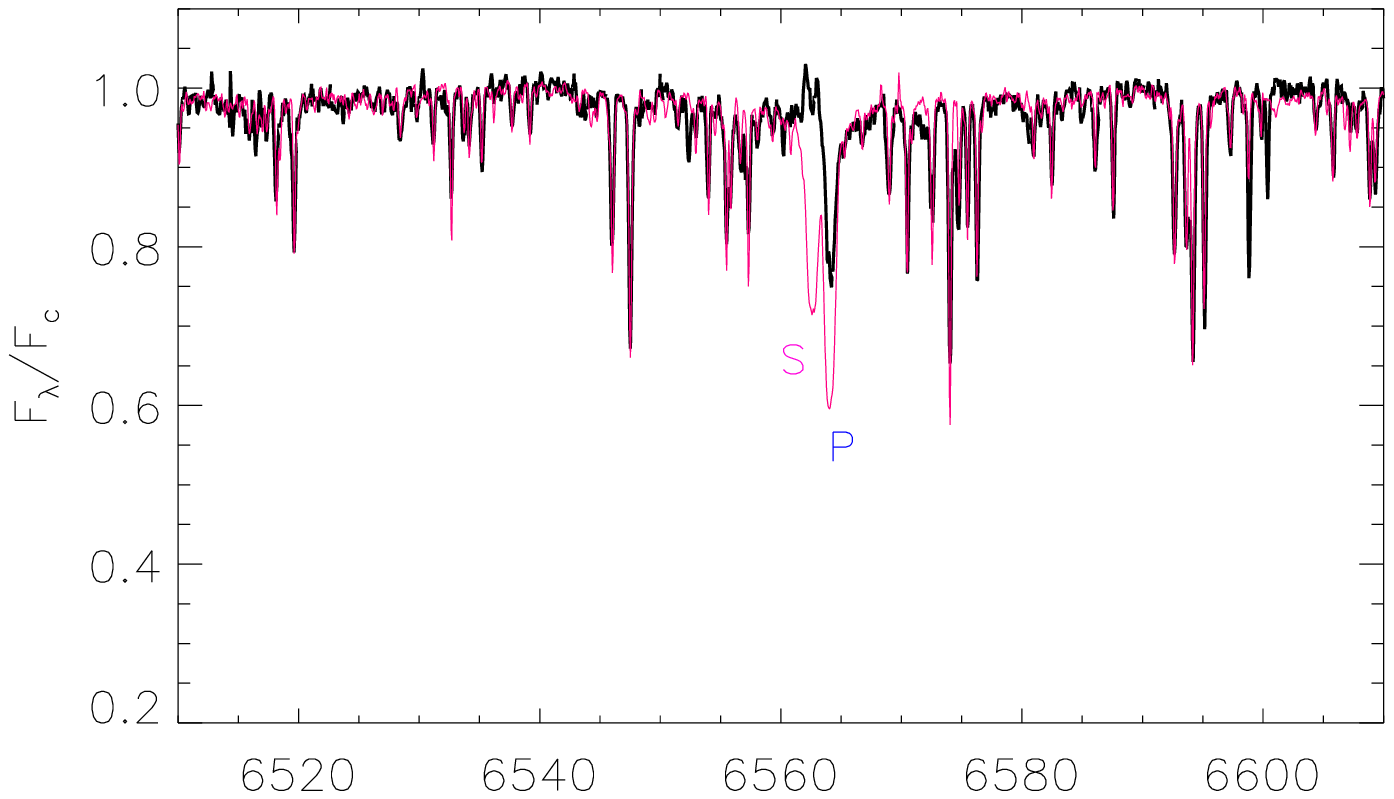}
\includegraphics[width=9.0cm,bb=25 10 453 260,clip]{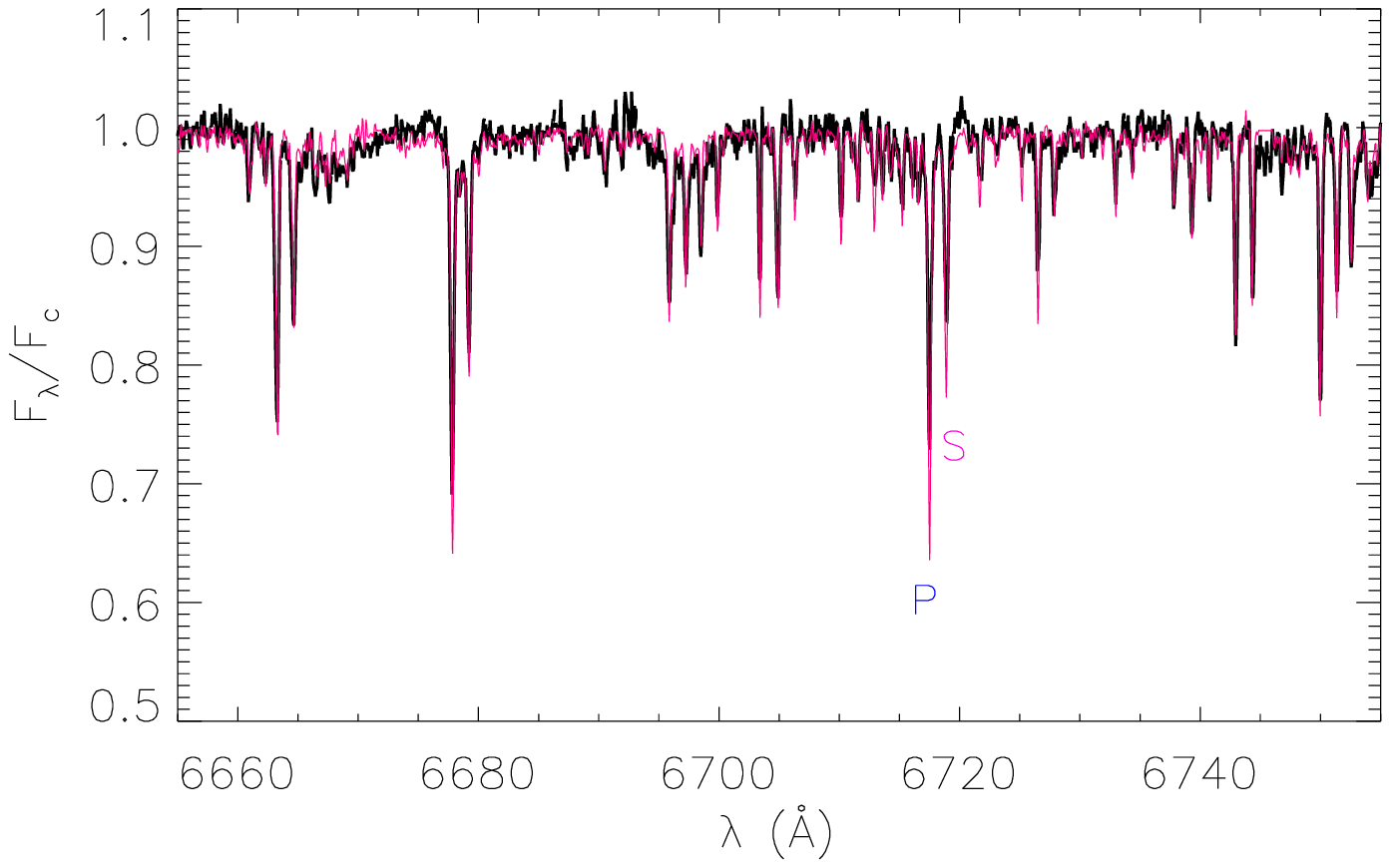}
\caption{Observed \textit{Aurelie} spectrum (black thick line) of the SB2 system BD+15~2915 (RasTyc~1547+1509, K2V+K3V) in the H$\alpha$ ({\it upper panel}) and lithium ({\it lower panel}) spectral regions, together with the synthetic spectrum built up with the weighted sum of two \textit{Elodie} reference spectra broadened at the $v\sin i$ of the system's components and Doppler-shifted according to its RV (thin red line). The H$\alpha$ cores of the primary (P) and secondary (S) components are marked in the upper panel, while the \ion{Ca}{i} $\lambda6717$ lines of the two components are labelled in the same way in the lower panel.} 
\label{Fig:Compo2}
\end{figure}
In the case of single stars or SB1 systems (i.e. only one significant minimum in the CCF; see Fig.~\ref{Fig:CCF} left panel), we run the ROTFIT code \citep{2003A&A...405..149F}, which finds the spectral type of the target searching into a library of standard-star spectra, the spectrum that gives the best match (minimum of the residuals) with the target one (see, for instance, \textit{online} Fig.~\ref{Fig:PTTSspec}), after rotational broadening (hence giving another independent measure of the star's $v\sin i$). As a standard-star library, we used a set of 185 spectra of stars with known astrophysical parameters \citep{2001A&A...369.1048P} retrieved from the ELODIE Archive \citep{2004PASP..116..693M}, which are well-distributed in effective temperature and gravity and in a suitable range of metallicities.

For double stars, we used another method (implemented as an IDL code), namely COMPO2 \citep{2006A&A...454..301F}, which searches for the best combination of two standard-star spectra able to reproduce the observed spectrum of the SB2 system. We give, as input parameters, the radial velocity and $v\sin i$ of the two components, which were derived from the CCF analysis. COMPO2 then finds, for the selected spectral region, the spectral types and fractional flux contributions that better reproduce the observed spectrum, i.e. which minimise the residuals in the collection of difference (observed\,$-$\,synthetic) spectra. For this task we used a smaller set of 87 reference spectra retrieved from the ELODIE Archive that are representative of stars with a nearly-solar metallicity, spectral types from early-F to early-M, and luminosity 
classes V, IV, and III.  An example of the results provided by COMPO2 is shown in Fig.~\ref{Fig:Compo2}.

For multiple systems, we made no attempt to derive astrophysical parameters, and these systems will be analysed in detail in a future dedicated work where a new COMPO code, designed for multiple systems, will be used.\\

The accuracy of the stellar parameters ($T_{\rm eff}$, $\log g$, [Fe/H]) depends not only on the S/N of the spectra but also on the homogeneous coverage of the grid of parameters by reference stars. To check the reliability of our determinations, we compared our results with those reported in the literature for six binaries classified with COMPO2. The determination of stellar parameters for the components of SB2 systems is a challenging task for any automatic spectral classification code, due to the blending of the spectral lines of the two stars. As COMPO2 is the equivalent of the ROTFIT code adapted to binary systems, we consider the accuracy on the parameters of SB2 components as an upper limit on the uncertainties for single stars. As apparent from Table~\ref{Tab:err}, our spectral type classification agrees with that from the literature within 1-2 subclasses for nearly all systems.
\begin{table*}[t]
\caption{Stellar parameters for 7 binaries with literature values.}
\begin{tabular}{lccccc}
\hline \hline
RasTyc Name & Name & Sp.Type$^{Our}$ &  Sp.Type$^{Lit}$ & $T_{\rm eff}^{Our}$ & $T_{\rm eff}^{Lit}$  \\ \hline
RasTyc~0702$-$0515 & AR Mon & K1 III / G2 V & ... / ... & 4395 / 5350 & 4500 / 5200$^{1}$  \\
RasTyc~0921$-$0640 & NY Hya & G2 V / G2 V & ... / ... & 5450 / 5450 & 5490 / ...$^{2}$  \\
RasTyc~1310+3556 & RS Cvn & F7 V / G8 V & F6 IV / G8 IV$^{3}$ & 6120 / 5420 & ... / ...  \\
RasTyc~1601+5120 & EV Dra & G7 V / K1 V & G8 V / K1 V$^{4}$ & 5400 / 5370 & ... / ...  \\
RasTyc~1825+1817 & AW Her & G2 V / K0 IV & G2 IV / K2 IV$^{5}$  & 5780 / 4870 & ... / ...  \\
RasTyc~1912+4619 & FL Lyr & G2 V / K1 V & F8 / G8$^{6}$ & 5730 / 5080 & 6080 / 5390$^{7}$ \\ 
\hline
\end{tabular}
\begin{list}{}{}
\item References. (1) \citet{2005AJ....129.2798W}; (2) \citet{2001A&A...374..980C}; (3) \citet{1990A&A...230..389S}; (4) \citet{1995AJ....110.2926H}; (5) \citet{2002A&A...387..850I}; (6) \citet{1986AJ.....91..383P}; (7) \citet{2007ActaAstronomica...57..301F}.
\end{list}
\label{Tab:err}
\end{table*}

\begin{figure*}[!ht]
\includegraphics[width=13.0cm,angle=90,bb=15 0 504 750,clip]{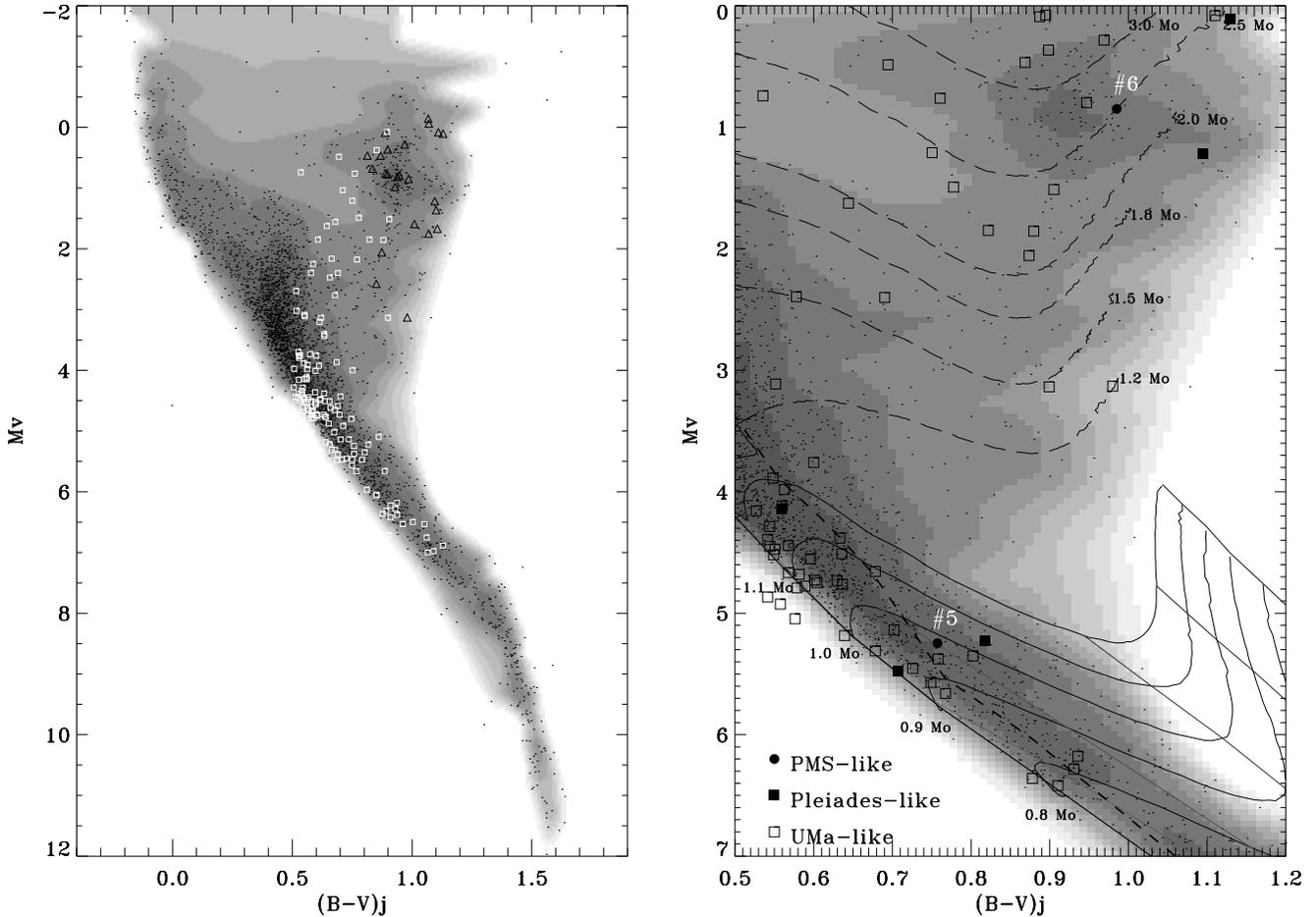}
\caption{{\it Left panel:} stars with trigonometric distance $\pi$ known better than 30\% (D30) and classified as class V (white square symbols) or class III, IV (black triangle symbols) superposed on an observational HR diagram of the $\approx$~3\,400 RasHip stars (dot symbols ; grey scale code their density gradient) with accurate distance  ($\sigma_{\pi}/\pi~\leqslant~$0.1) and colour ($\sigma_{B-V}~\leqslant~$0.025). \textit{Right panel} : Zoom on the late-type region of the HR diagram showing the main sequence and giant clump regions together with D30 stars classified as \textit{PMS-like} (filled circle ; \#i show cross-match with Table~\ref{Tab:MVCand}), \textit{Pleiades-like} (filled square) and \textit{UMa-like} (open square) based on their lithium abundance. Siess et al. (1997) PMS (from 0.8 to 1.2\,M$_{\odot}$) and post MS (from 1.2 to 3.0\,M$_{\odot}$) evolutionary tracks and isochrones (1, 3, 10, 30, and 50\,Myr from top to bottom) have been overlaid. Zero age main sequence (bold solid line) and terminal age main sequence (dashed bold line) are shown as well. The 2 \textit{PMS-like} stars are discussed in detail in Sect.~\ref{Sec:MG}.}
\label{Fig:HRdiag}
\end{figure*}

As a second test, we selected stars classified by ROTFIT for which the trigonometric distance has an uncertainty below 30\% (D30) and plotted them on an HR diagram, as illustrated in Fig.~\ref{Fig:HRdiag} left panel. It readily shows that none of the stars lying on the MS was misclassified as a giant and that, among the 50 stars with $B-V>0.8$, only 6 (i.e. 12~\%) giants or subgiants (according to their HR diagram location) were misclassified as dwarfs. On the other hand, stars leaving the upper main sequence, but which have not yet reached the giant clump, are systematically classified wrongly as dwarfs by our automatic procedure. Quantitatively, 22 among the 111 stars with $B-V<0.8$, i.e. $\backsimeq$~20\%, are misclassified. This is likely the result of the lack of similar stars in the reference library (due to the very short lifetime of this evolutionary phase), but not because of the formal algorithm of our code.\\

Thus, we consider our spectral classification affected by an uncertainty that has a typical value of one spectral subclass ($\Delta T_{\rm eff}\approx 150$\,K) but can reach two spectral subclasses ($\Delta T_{\rm eff}\approx300$\,K) in the less favourable cases (especially for SB2 systems). We also stress that $\backsimeq$~20\% of our stars bluer than $B-V=0.8$ can suffer luminosity misclassification; i.e., they could be stars on their way towards the giant branch that have a lower gravity compared to the MS stars with the same temperature.\\
\begin{table*}
\caption{Consistency of the stellar parameters derived from the \textit{Aurelie} (left) and \textit{Elodie} (right) spectrographs.}
\begin{tabular}{lcccccccccccc}
\hline \hline
RasTyc Name & \multicolumn{2}{c}{RV (km~s$^{-1}$)} & \multicolumn{2}{c}{$v\sin i$ (km~s$^{-1}$)} & \multicolumn{2}{c}{$T_{\rm eff}$ (K)} & \multicolumn{2}{c}{$\log g$} & \multicolumn{2}{c}{$[Fe/H]$} \\ 
\hline
RasTyc~2100+4530 & $-23.3$ &  $-23.1$ & 6   &  6  & 5720  &  5730 & 4.23  &  4.42 & $-0.01$  &  $ 0.01$ \\
RasTyc~2114+3941 & $-10.8$ &  $-11.2$ & 24  &  27 & 5730  &  5730 & 4.14  &  4.28 & $-0.23$  &  $-0.53$ \\
RasTyc~2118+2613 & $-16.3$ &  $-16.8$ & 7   &  6  & 5230  &  5330 & 4.31  &  4.52 & $-0.09$  &  $-0.10$ \\
RasTyc~2132+3605 & $ -6.2$ &  $ -6.8$ & 5   &  4  & 5140  &  5000 & 3.20  &  2.84 & $-0.03$  &  $-0.05$ \\
RasTyc~2137+1946 & $ -5.2$ &  $ -6.0$ & 4   &  3  & 4900  &  4870 & 3.03  &  2.85 & $ 0.04$  &  $-0.01$ \\
\hline
\end{tabular}
\label{Tab:param}
\end{table*}

The data acquired from the various telescopes/spectrographs were treated in slightly different ways. That is why a few single stars were observed with different instruments for checking the consistency of our results. As apparent from Table~\ref{Tab:param}, which compares parameters derived from the T1.52$+$\textit{Aurelie} and T1.93$+$\textit{Elodie} spectrographs, there is no discrepancy or systematic effects and the agreement is very good on average. Consistency of measurements from OHP and OAC spectra has already been demonstrated in \cite{2006A&A...454..301F} and \cite{2008A&A...490..737K}.

\section{Chromospheric emission and lithium abundance}
\label{Sec:ActAg}
\subsection{Chromospheric emission}
\label{Sec:Chromo}
In addition to the fundamental parameters (effective temperature, gravity, metallicity, and $v\sin i$), the ROTFIT and COMPO2 codes allowed us to derive stellar activity and age. We estimated the chromospheric emission level with the ``spectral subtraction'' technique  \citep[see, e.g.,][]{1985ApJ...289..269H, 1985ApJ...295..162B,1994A&A...284..883F, 1995A&AS..109..135M}, which is based on the subtraction of a reference ``non-active'' spectrum built up with synthetic model-atmosphere spectra or with observed spectra of slowly-rotating stars of the same spectral type with a negligible level of chromospheric activity. The net equivalent width of the H$\alpha$ line, $W_{\rm H\alpha}^{em}$, was measured in the spectrum  obtained after subtracting the non-active template by integrating the residual H$\alpha$ emission profile. Although it is not easy to get truly chromospherically non-active stars, especially for the K-M spectral types, using the spectra exhibiting the minimum H$\alpha$ residual flux as ``non-active'' templates allowed us to pick up stars with a  moderate activity and to evaluate its level.
The H$\alpha$ luminosity, which is a parameter more indicative of the chromospheric activity level, was calculated from the net H$\alpha$ equivalent width, the distance $d$, and the continuum Earth flux at the H$\alpha$ wavelength, $f_{6563}$, according to the following equation:
\begin{eqnarray}
L_{\rm H\alpha} & = & 4\pi d^2 f_{6563}W_{\rm H\alpha}^{em} \nonumber\\
                & = & 4\pi d^2 \frac{F_{6563}}{F_{5556}}10^{(-0.4V_0-8.451)}W_{\rm H\alpha}^{em}, 
\end{eqnarray}
{\noindent where $10^{(-0.4V_0-8.451)}$ is the Earth flux at 5556\,\AA\  of a star with de-reddened magnitude $V_0$ \citep{1992Sci...257.1978G} and the continuum flux-ratio $\frac{F_{6563}}{F_{5556}}$ was evaluated from NextGen synthetic low-resolution spectra \citep{1999ApJ...512..377H} at the temperature found with the ROTFIT code.}

\subsection{Lithium abundance}
\label{Sec:Lithium}
Lithium is strongly depleted when mixing mechanisms pull it deeply into the convective layers of late-type stars. A strong \ion{Li}{i}\,$\lambda$6707.8 photospheric absorption line is thus generally considered as a youth indicator \citep[e.g.,][]{1998AJ....116..396S}. We measured the equivalent width of the \ion{Li}{i}\,$\lambda$6707.8 line on the residual spectrum, once it hs been checked that this feature is not present in the spectrum of the reference star. The subtraction technique  allowed us to remove the contribution of the small nearby \ion{Fe}{i}\,$\lambda$6707.44 line, whose intensity is expected to be nearly the same in the reference spectrum that best represents the target. We estimate the lower limit for positive detection of  the \ion{Li}{i}\,$\lambda$6707.8 line to be approximately 10~m$\AA$, although it strongly depends on the S/N of the spectrum. The lithium abundance was finally deduced from the \ion{Li}{i}\,$\lambda$6707.8 line equivalent width (EW(Li)) and effective temperature by linear interpolation of the values tabulated by \citet{1996A&A...311..961P}.

\section{Discussion}
\label{Sec:Disc}

Although a detailed multidimensional analysis of all parameters derived from our spectra is beyond the scope of this paper, in the following subsections we briefly discuss the fraction of multiple systems (Sect.~\ref{Sec:Mult}), as well as the evolutionary status (Sect.~\ref{Sec:Evo}) and age classification (Sect.~\ref{Sec:Age}) of our star sample. Finally we investigate in detail the lithium-rich stars (Sect.~\ref{Sec:LiG} and Sect.~\ref{Sec:MG}) of our sample, among which five appear as possible members of young stellar kinematical groups in the solar neighbourhood.

\subsection{Binaries and multiple systems}
\label{Sec:Mult}

\citet{1996A&A...307..121B} have pointed out that ROSAT unresolved binaries are X-ray sources statistically brighter than single stars. Thus, X-ray flux-limited surveys suffer from a detection bias towards binarity. Indeed, a large number of spectroscopic binaries (SB) have been discovered in this sample. More quantitatively, if restricting ourselves to stars observed with \textit{Aur\'elie} in both spectral regions, our sample contains 210 single stars, 47 single-lined spectroscopic binaries (SB1), 38 double-lined spectroscopic binaries (SB2), and 9 bona-fide triple systems, among which 2 are  potential multiple (more than 3 components) systems. Most of them are new detections (according to the SIMBAD database). Thus, in our sample we have percentages of 15.5\%, 12.5\%, and 3.0\% for SB1, SB2, and multiple systems, respectively. Altogether, non-single stars account for 31\% of the sample.\\

Six binaries (3 SB1 and 3 SB2; \citealt{2006A&A...454..301F}) as well as 3 triples systems \citep{2008A&A...490..737K} have already been studied in great detail. The high-quality RV curves have been solved for these systems, and the orbital and physical parameters obtained. Although no significant conclusions can be drawn with respect to evolutionary status of SB, we are continuing to monitor a representative sample of RasTyc SB.
\begin{figure}[]
\includegraphics[width=9.0cm]{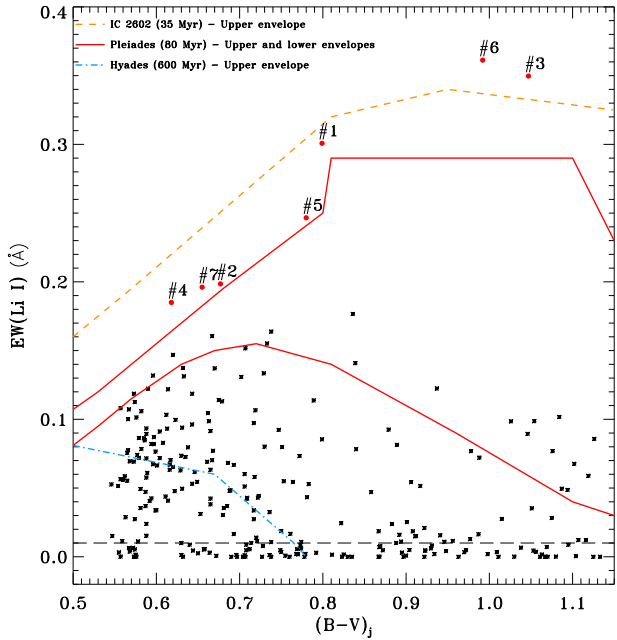}
\caption{\ion{Li}{i}\,$\lambda$6707.8 line equivalent widths versus $B-V$ colour index for stars of our sample. The lines represent the boundaries for IC~2602 (orange dash), Pleiades (red solid), and Hyades (blue dash-dot) clusters. The black long-dashed line shows our lower limit for positive detection of the  \ion{Li}{i}\,$\lambda$6707.8 line. Seven stars (red dots next to \#i symbols; see Table~\ref{Tab:MVCand} for cross-match) show lithium abundance in excess of the lithium-richest Pleiades stars and 2 are located above IC~2602 lithium upper envelope.}
\label{Fig:Li}
\end{figure}

\subsection{Evolutionary status}
\label{Sec:Evo}

Stellar soft X-ray surveys have long been assumed to be mostly composed of young active late-type stars with a possible contamination by older active binaries (i.e. RS CVn systems; \citealt{1995A&A...296..370S}). However, because of the lack of knowledge on this population (density and scale height), the comparison of stellar X-ray population models with optically-identified X-ray sources generally assumed that the contamination by evolved systems is at most only marginal. Depending on its strength, the impact of the contamination by older systems that mimic young stars can lead to significant error on the derived recent local star formation rate \citep{2008A&A...483..801A}. Assuming that the RasTyc sample is representative of X-ray unbiased and magnitude limited surveys, we derived an overall contamination by evolved stars (likely RS CVn systems) of about 35~\% with a peak of 60~\% in the K-type stars region ($0.8 \lesssim B - V \lesssim 1.1$), whereas this fraction decreases to only 20~\% and 10~\% for hotter (F-G) and cooler (M) stars, respectively.  

\subsection{Age}
\label{Sec:Age}

The lithium abundance to indicate the empirical age is now commonly accepted for stars cooler than $B-V\simeq 0\fm6$. It is possible to define with high confidence in a colour versus \ion{Li}{i}\,$\lambda$6707.8 line equivalent width diagram the loci occupied by \textit{PMS-like}, \textit{Pleiades-like}, \textit{UMa-like}, \textit{Hyades-like}, and \textit{Old} (i.e. age greater than one Gyr) stars. Such a diagram is presented in Fig.~\ref{Fig:Li} for only our single and SB1 stars, together with the IC~2602 \citep{2001A&A...379..976M}, Pleiades \citep{1993AJ....106.1059S, 1997Sci...276.1363N}, and Hyades \citep{1990AJ.....99..595S} envelopes. Some of our stars exhibit a lithium content larger than or compatible with that of IC~2602, indicating PMS or PTTS candidates (see discussion in Sect.~\ref{Sec:MG}). Other stars occupy a domain similar to that of the Pleiades, UMa, or Hyades, so they likely have similar ages. There are also stars with no photospheric \ion{Li}{i}\,$\lambda$6707.8 line detected or with lithium content below the Hyades lower envelope, which are likely to be slightly older than one giga-year according to current stellar X-ray population models.\\

The fractions of \textit{PMS-like}, \textit{Pleiades-like}, \textit{UMa-like}, \textit{Hyades-like}, and \textit{Old} stars are 2.6\%, 7.0\%, 39.1\%, 19.2\%, and 32.1\%, respectively. Interestingly, the fraction of \textit{PMS-like} stars derived from the only lithium criterion is exactly the one predicted by \citet{1999A&A...351.1003G} from a statistical analysis of RasTyc/RasHip samples. Altogether, the \textit{very young} stars (i.e. \textit{PMS-like} plus \textit{Pleiades-like} and \textit{UMa-like}) contribute about half of the sample. If we consider all the \textit{young stars} (i.e. \textit{very young} plus  \textit{Hyades-like}, namely stars younger than about one giga-year), this fraction increases to about 70\%. Outside star formation sites (clusters or associations), young stars generally contribute only marginally ($\leqslant$~10\% assuming a constant star formation rate) to optical star counts, in contrast to the high fraction found by us. Although lithium is depleted very rapidly in the convective envelopes of cool stars, it is not surprising to detect a significant fraction of lithium-rich stars in our sample, which is heavily biased towards young stars thanks to X-ray selection.\\

A crude analysis of the galactic latitude distribution of \textit{young} and \textit{old} populations of RasTyc stars does not reveal any enhancement of \textit{young} star number towards the galactic plane. This is actually not surprising because the majority of our stars (70\%) are located within 100\,pc of the Sun, a radius that is smaller than or comparable to the scale height of young stars estimated in between 70~pc (assuming the scale height of molecular clouds ; \citealt{1984ApJ...276..182S}) and 140~pc (as computed in the Besan\c{c}on synthesis model ; \citealt{1986A&A...157...71R, 2003A&A...409..523R}). The comparison of our sample with the predictions of a stellar X-ray population model \citep{1996A&A...316...89G} will be the scope of a dedicated paper.\\

\subsection{Lithium-rich giants}
\label{Sec:LiG}

In the right panel of Fig.~\ref{Fig:HRdiag}, we plot a zoom on the late-type region of the HR diagram showing the main sequence and giant clump regions, together with D30 stars classified as \textit{PMS-like}, \textit{Pleiades-like}, and \textit{UMa-like} based on their lithium abundance. It shows that, unforeseeasly, a significant fraction (23 out of 50, i.e. 46~\%) of the evolved stars are also lithium-rich. Most of them are classified as \textit{UMa-like} but a few of them display an even larger lithium content and are classified as \textit{Pleiades-like} or \textit{PMS-like}, accordingly. These stars have early-F or A-type progenitors on the main sequence ($M~\simeq$~1.2-3.5~$M_{\sun}$), which are known to be very fast rotators but which generally do {\em not} display any \ion{Li}{i}\,$\lambda$6707.8 absorption line feature,
although some Ap stars have lithium-rich spots \citep{2004IAUS..224..692D, 2005IAUS..228...89D}.\\

\begin{table*}[t]
\caption{PMS-like candidates based on the lithium criteria only.}
\small
\begin{tabular}{llccccccccc}
\hline \hline
\# Name & RasTyc Name & $\alpha$ (2000) &  $\delta$ (2000) & PSPC  & Sp. T. & $T_{\rm eff}$ & $\log g$ & [Fe/H] & $\log$(n$_{LiI}$) & $\log$(H$\alpha$) \\
     &              & h m s & $\circ$ ' '' & (ct/s) & L. cl. & (K) & & & & \\ 
\hline
\#1 BD+45~598 & RasTyc~0221+4600 &  02 21 13.0  & +46 00  7.6 &      0.35 &  K1 V &  5149 &       4.04 &    -0.09 &  3.20 & 28.97 \\
\#2 BD+29~525 & RasTyc~0307+3020 &  03 07 59.2 & +30 20 26.7 &  0.45 &  G2 V &   5827 &       4.31 &    -0.03 & 3.24 & ... \\
\#3 HD~275148$^{*}$ & RasTyc~0319+4212 &  03 19 52.8  & +42 12 31.1 &     0.03 &  G7 III &   5158 &       3.39 &      -0.00 & 3.66 & 30.46 \\
\#4 HD~22179 & RasTyc~0335+3113 &  03 35 29.9  & +31 13 37.8 &      0.26 &  G3 V &    5705 &       4.32 &    -0.06 & 3.13 & 29.37 \\
\#5 HD~23524 & RasTyc~0348+5202 &  03 48 23.0  & +52 02 16.9 &      0.86 &  G9 V &       5213 &       4.22 &    -0.06 & 2.94 & 29.33 \\
\#6 HD~170527$^{*}$ & RasTyc~1825+6450 &  18 25 10.1  & +64 50 17.8 &      0.23 &  G4 III &       5118 &       3.60 &    -0.09 & 3.69 & 31.10 \\
\#7 BD+44~3670 & RasTyc~2100+4530 &  21 00 47.0  & +45 30 10.4 &      0.35 &  G2 V &       5719 &       4.23 &   -0.01 & 3.21 & 29.26 \\
\hline
\end{tabular}
\normalsize
\begin{list}{}{}
\item[$ ^{*}$] were later discarded as PMS candidates because of spectral classification and kinematic criteria inconsistancies.
\end{list}
\label{Tab:LiRich}
\end{table*}

In a series of works, de la Reza and co-workers (\citealt{1996ApJ...456L.115D, 1997Ap&SS.255..251D, 1997ApJ...482L..77D, 2000ApJ...535L.115D}; \citealt{2002AJ....123.2703D}; \citealt{2006cams.book..196D}) proved that {\it all} K-giant stars, with masses between 1 and 2.5 solar masses, become lithium-rich during the red giant branch stage. For these authors, these stars are normal giants going through a short period of lithium enrichment. This episode, which could be cyclic, would take origin in the changes of the stellar structure of the star. As these stars evolve towards the blue part of the clump (as helium-burning giants), they develop outer convection zones and consequently a high magnetic activity level, because of the convection and fast rotation \citep{1993ApJ...403..708F}, which yields the X-ray emission.\\ 

The mechanism leading to simultaneously high rotation, high lithium abundance, and enhanced activity is not yet understood. A first scenario was proposed by \cite{1971ApJ...164..111C} for the asymptotic giant branch evolution phase: the convection zone can go so deep inside the stellar interior that it carries material containing $^{7}$Be to upper cooler layers where it decays into $^{7}$Li.
\cite{1993ApJ...403..708F} have proposed a scenario in which internal convective motions penetrate into the rapidly rotating core dredging up to the surface high angular momentum material and lithium recently synthesised. In contrast, \cite{1996ApJ...456L.115D} suggests that the enrichment of lithium is an obliged and rapid phase for all the K-type giants, produced by a mechanism similar to the one invoked by \cite{1971ApJ...164..111C}, which leads to the formation of a shell of gas and dust.
Other studies suggest that the lithium enrichment could originate in the ingestion of substellar bodies like planets or brown dwarfs (\citealt{1999MNRAS.304..925S}, \citealt{2000A&A...358L..49D}).
At this time, the \ion{Li}{i} line should appear in their spectra. However, the mechanism of lithium formation loses its efficiency after some 10$^{5}$ years so that thereafter the lithium is progressively destroyed in the deeper layers where it is brought by the convecting mixing. The observation of a strong \ion{Li}{i}\,$\lambda$6707.8 absorption line would indicate that these stars have just reached this evolutionary phase without having enough time to deplete the photospheric lithium.\\

Last but not the least could be lithium preservation in atmospheric stellar layers after some process like those acting in Ap stars (high and closed magnetic fields, diffusion, etc.) when the stars evolve out of the main sequence; but this scenario should suggest that lithium-rich K-giants may have spotted lithium-rich Ap progenitors, a new scenario that needs more detailed analysis in the future.\\

The discovery of this large number of lithium-rich giants adds more than 40\,\% to the known ones (40 quoted by \citealt{1997ApJ...482L..77D}, 15 quoted by \citealt{2006A&A...449..211P} and some other dispersed in the literature). Our results reinforce those of \cite{1994ASPC...64..279F}, who found that 58\,\% of single {\em active} giants have high lithium abundance (36\,\% for the giants in binary systems) and are also in line with a 50\,\% of lithium-rich stars among rapidly-rotating K-giants found by \cite{2002AJ....123.2703D}. The emerging picture is that a high or moderate atmospheric lithium abundance, coronal activity, and a rather rapid rotation are characteristic of an evolutionary phase that should be  relatively short, given the small fraction (1-2\,\%) of lithium-rich stars among the whole sample of G- and K-type giants (see \citealt{2000A&A...359..563C}).

\subsection{Moving groups candidates}
\label{Sec:MG}
\begin{figure*}[!t]
\includegraphics[width=18.0cm]{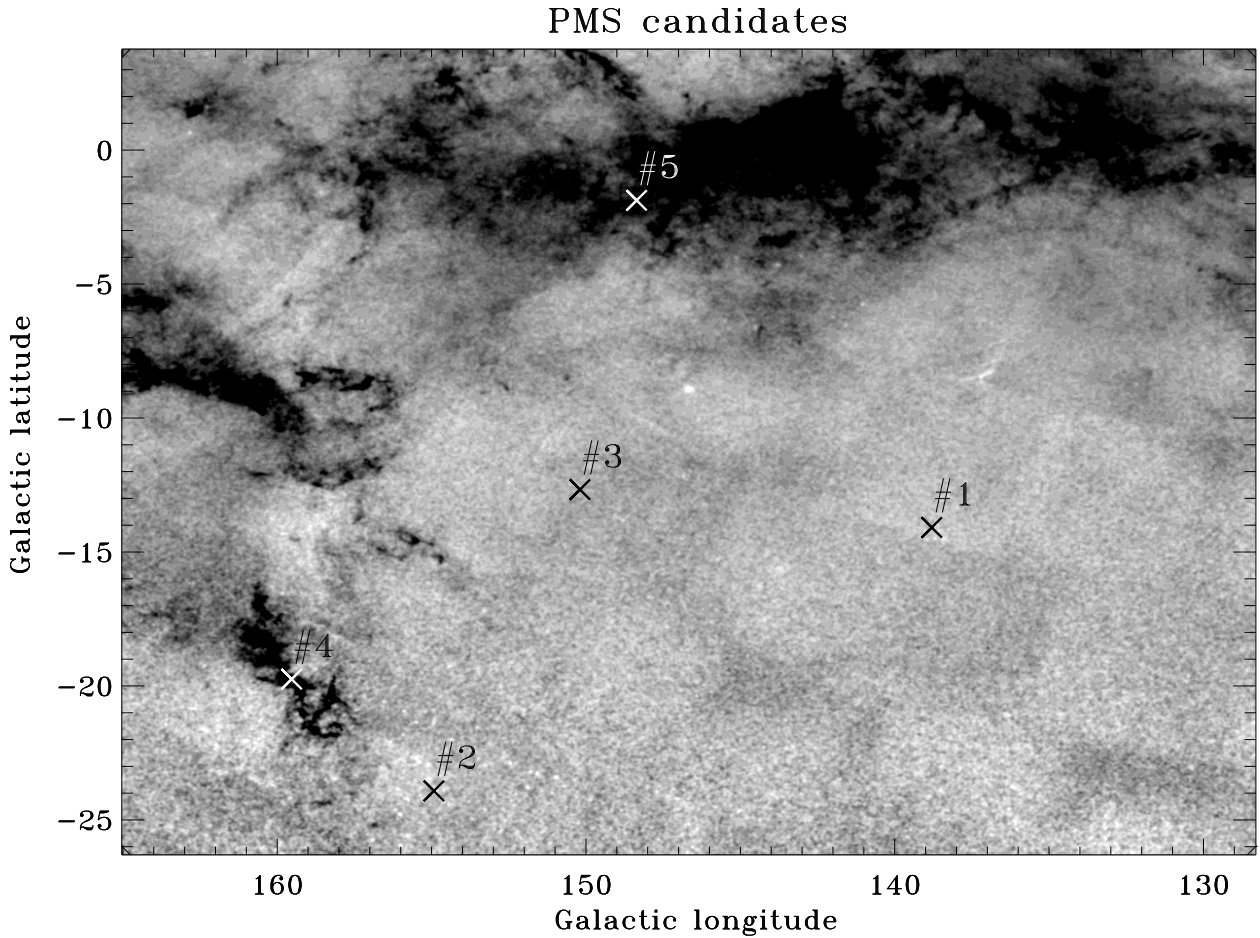}
\caption{Position (galactic coordinates) of PMS-like candidates overplotted on a map of dark clouds enhancing interstellar material (see \#i next to cross symbols for cross match with Table~\ref{Tab:MVCand})}
\label{Fig:PMS_FC}
\end{figure*}

The highest EW(Li) values measured in our sample reach over 350~m$\AA$ for 2 stars, namely RasTyc~1825+6450 (\#6) and RasTyc~0319+4212 (\#3). Together with 5 other stars (RasTyc~0221+4600, \#1; RasTyc~0307+3020, \#2; RasTyc~0335+3113, \#4; RasTyc~0348+5202, \#5 and RasTyc~2100+4530, \#7) showing EW(Li) compatible with that of IC~2602, they have a lithium content close to the primordial abundance $\log$(n$_{LiI}$)~$\sim$~3.2 (see Table~\ref{Tab:LiRich}) and can be considered as trustworthy PMS candidates. The observed spectra of our 7 candidates are presented in Fig.~\ref{Fig:PTTSspec} and Fig.~\ref{Fig:LiRichGiantspec}. It readily shows that, although all sources (with the only exception of RasTyc~0319+4212) display significantly filled-in H$\alpha$ line profiles typical of stellar activity, none of them exhibits strong emission features above the continuum that are distinctive of classical T Tauri stars. On the other hand, no spectroscopic criterion has yet been clearly identified to disentangle weak-line T Tauri stars (WTTS) from PTTS. \citet{1998AJ....115..351M} has tentatively established a divided line in the $EW(Li)~-~T_{\rm eff}$ diagram for distinguishing between WTTS and PTTS but unfortunately our candidates fall in the region of confusion where uncertainties prevent us from distinguishing them.\\

Although RasTyc~1825+6450 and RasTyc~0319+4212 are the lithium-richest stars of our sample, they were classified as giants by our automatic spectral classification code (see Sect.~\ref{Sec:TeffloggFeH}). A strong lithium absorption in the atmosphere of giants stars is not an obvious indicator of youth, as we shortly discussed in Sect.~\ref{Sec:Age}. However, very young PMS stars that are still in the radiative part of their pre-main sequence evolutionary track can mimic evolved (subgiant or giant) stars with typical gravities in the range $\log~g$~=~3.0-3.5, so our code can find a better match with a subgiant or giant template. In addition, their position in the HR diagram lies in the region occupied by subgiants of higher mass. Thus, we cannot reject them from the list of PMS star candidates.\\

The positions (galactic coordinates) of candidates \#1, \#2, \#3, \#4, and \#5 are overplotted in Fig.~\ref{Fig:PMS_FC} on a map of dark clouds published by \citet{2005PASJ...57S...1D}. The closest prominent star forming region is the Taurus-Auriga complex located more than 10 degrees east of star \#4 outside the portion of the displayed map. Although this candidate is projected directly in front of the Perseus molecular cloud, in the vicinity of the Per OB2 association \citep{1952BAN....11..405B} and IC~348 young star cluster, its distance is hardly compatible with these background objects, even assuming that the star is overluminous with respect to MS calibration because of its PMS nature. Thus, we consider this star as unrelated to the Perseus star-forming region. Star \#5 is projected in front of unrelated background dark clouds on the galactic plane, while stars \#1, \#2, and \#3 are located below the galactic plane in regions free from dense interstellar matter. Star \#7 (not shown) is located just north of the unrelated North America nebula, while star \#6 (not shown) lies above the galactic plane in the Draco constellation. Since none of them is located towards any close (i.e. within 200~pc) prominent star-forming region, they appear to be likely members of young stellar kinematic groups (SKG) in the solar neighbourhood. \\

\begin{figure*}[!t]
\includegraphics[width=9.0cm]{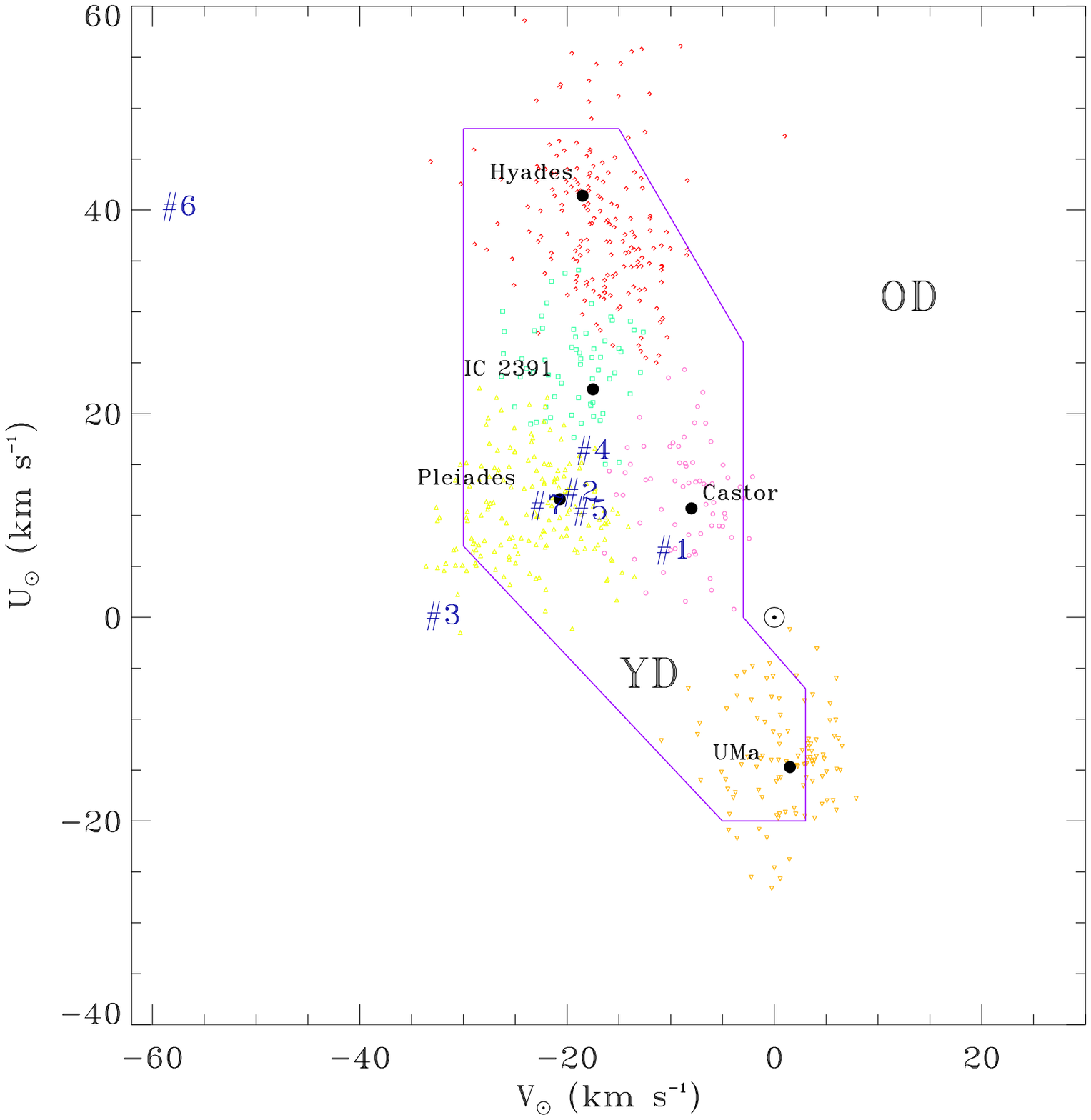}
\includegraphics[width=9.0cm]{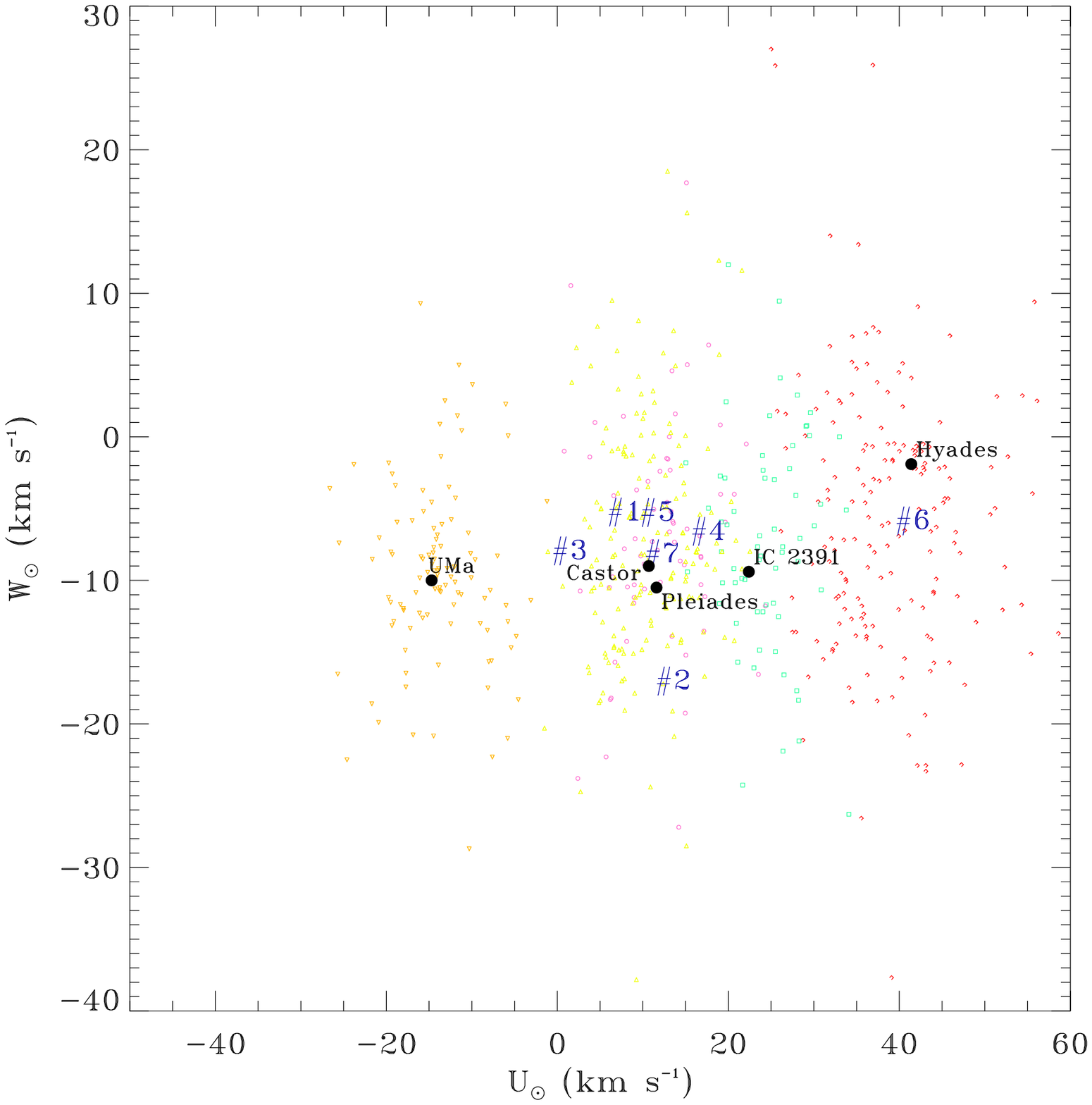}
\caption{The U$-$V (\textit{left panel}) and U$-$W (\textit{right panel}) kinematic diagrams for our moving group candidates (\#i symbols ; see Table~\ref{Tab:MVCand} for cross-match). The average velocity components (dots) of some young SKG and those of some late-type stars members of these young SKG are also plotted (square, triangle, circle, upside down triangle, and U symbols for the IC~2391 supercluster, Pleiades, Castor, UMa moving groups, and Hyades supercluster, respectively). The locus of the young-disc (YD) and old-disc (OD) populations are also marked.}
\label{Fig:UVW_PMS}
\end{figure*}

To assess possible membership as already known SKG, we computed the space velocity components (U, V, W; see Table~\ref{Tab:MVCand}) for these seven stars in a lefthanded coordinate system using Hipparcos/Tycho parallaxes (\#5 and \#6), otherwise photometric distance derived from absolute magnitude calibration from \citet{2000S&T...100a..72C} and Tycho2 \citep{2000A&A...355L..27H} proper motions together with our measured radial velocities.
The U$-$V and U$-$W kinematic diagrams are shown in Fig.~\ref{Fig:UVW_PMS} for stars listed in Table~\ref{Tab:LiRich}. We also show the location of stars satisfying one of the two Eggen criteria for the major stellar kinematics groups discussed in \citet{2001MNRAS.328...45M}, namely the IC~2391 supercluster ($\sim 50$~Myr), the Pleiades SKG ($\sim 100$~Myr), the Castor SKG ($\sim 200$~Myr), the Ursa Major (UMa) group ($\sim 300$~Myr), and the Hyades supercluster ($\sim 600$~Myr). The locus of the young-disc (YD; age~$\leqslant 2$~Gyr) and the Sun position are shown as well. 

Figure~\ref{Fig:UVW_PMS} clearly shows that five stars have U, V, and W compatible with the IC~2391 and/or Pleiades and/or Castor kinematics, which are the youngest moving groups considered in this study. RasTyc~0319+4212 (\#3) is at most marginally compatible with the Pleiades SKG  kinematics while RasTyc~1825+6450 (\#6) is located far outside of the YD locus in the U$-$V and V$-$W planes and displays kinematics inconsistent with any SKG. Amazingly, these two stars are the ones classified as giants by our automatic spectral classification programme.

The membership in SKG can be analysed more objectively using probabilistic methods (\citealt{2008PhDT.........7V}; Klutsch et al. 2009, in preparation). We computed the probability that a star with heliocentric velocities U, V, and W has a galactic motion compatible with one of the five aforementioned comoving groups of stars. Results are presented in Table~\ref{Tab:MVCand}. We accepted a star as a possible new member of an SKG and flagged it with a \textit{Y} in Table~\ref{Tab:MVCand} if the probability (bracketed near the acceptance flag) exceeded 10\%. According to this criterion, we find 5 stars displaying galactic motion compatible with the IC~2391 and/or Pleiades and/or Castor SKG. Two of them (RasTyc~0348+5202 and RasTyc~2100+4530) have probability exceeding 80\% which leaves little doubt as their membership to the Pleiades SKG. With probability higher than 50\%, RasTyc~0307+3020 is another highly probable new Pleiades SKG member. The results are more ambiguous for RasTyc~0335+3113, which is either a new IC~2391 or Pleiades SKG member with probability close to 30 and 50~\%, respectively. RasTyc~0221+4600 is our only candidate probably associated with the Castor SKG, although again we cannot rule out its connection to the Pleiades SKG. Unsurprisingly, the quantitative analysis confirmed that none of the stars classified as giants (RasTyc~0319+4212 and RasTyc~1825+6450) are associated with any of the SKG.
\begin{table*}[]
\caption{Kinematics data and membership probability of moving group candidates.}
\begin{tabular} {llllllllll}
\hline \hline
\# Name & RasTyc Name & U$_{\sun}$ & V$_{\sun}$ & W$_{\sun}$ & IC~2391 & Pleiades & Castor & UMa & Hyades \\ 
 & & (km~s$^{-1}$) & (km~s$^{-1}$) & (km~s$^{-1}$) &  & & & & \\
\hline
\#1 BD+45~598 & RasTyc~0221+4600 & 6.7 & -10.7 & -5.3 & N (0.0) & Y (17.6) & \textbf{Y} (65.4) & N (0.0) & N (0.0) \\
\#2 BD+29~525 & RasTyc~0307+3020 & 12.3 & -19.6 & -17.0 & N (0.9) & \textbf{Y} (54.1) & N (0.9) & N (0.0) & N (0.0) \\
\#3 HD~275148 & RasTyc~0319+4212 & 0.2  & -32.9 & -8.0 & N (0.0) & N (9.8) & N (0.0) & N (0.0) & N (0.0) \\
\#4 HD~22179 & RasTyc~0335+3113 & 16.4 & -18.4 & -6.6 & Y (28.5) & \textbf{Y} (48.5) & N (6.5) & N (0.0) & N (1.2) \\
\#5 HD~23524 & RasTyc~0348+5202 & 10.5 & -18.7 & -5.3 & N (0.6) & \textbf{Y} (82.1) & N (6.6) & N (0.0) & N (0.0) \\
\#6 HD~170527 & RasTyc~1825+6450 & 40.4 & -58.3 & -5.9 & N (0.0) & N (0.0) & N (0.0) & N (0.0) & N (0.0) \\
\#7 BD+44~3670 & RasTyc~2100+4530 & 11.0 & -22.8 & -8.2 & N (0.8) & \textbf{Y} (99.3) & N (0.6) & N (0.0) & N (0.0) \\
\hline
\end{tabular}
\label{Tab:MVCand}
\end{table*}

The very good Hipparcos parallaxes ($\sigma_{\pi}/\pi \leqslant 0.1$) of RasTyc~0348+5202 (\#5) and RasTyc~1825+6450 (\#6) allowed their positions to be plotted  in the HR diagram, as illustrated in the right panel of Fig.~\ref{Fig:HRdiag}. As expected, RasTyc~1825+6450 lies in the giant clump and is compatible with a 2.5 solar mass progenitor $\sim$~500~Myr old according to post evolutionary tracks. On the other hand, RasTyc~0348+5202 is compatible with a $\sim$~30~Myr old solar mass PTTS evolving towards the ZAMS. These two particular cases for which we can access their evolutionary status, based on independent methods, give us a high confidence on our analysis and results.

\section{Conclusions}
\label{Sec:C&P}

In this paper we have presented the first results for a high-resolution spectroscopic survey of optical counterparts of X-ray sources  in the northern hemisphere aiming at a thorough characterisation of the presumably youngest field stars in the solar neighbourhood. In particular, we focused on the $\backsim$~400 optically brightest candidates for which we acquired about 800 spectra in the  H$\alpha$ and/or lithium spectral regions. Using properly developed methods, we determined accurate radial and rotational velocities, effective temperature, and gravity. Moreover, an evaluation of the age (based on the \ion{Li}{i}\,$\lambda$6707.8 line depth) and of the chromospheric activity level were performed. We have also identified several new binaries and multiple systems. We discussed the fraction of these systems in stellar soft X-ray surveys and showed that the contamination by lithium-rich evolved stars is not negligible. This sample of optically-bright and X-ray flux-limited sources turns out to mainly be composed of young stars ($age <$\,1 Gyr) with a smaller contribution from an older population that was probably born within the last giga-years. Outstanding results are the discovery of about 50 new lithium-rich giants and new members of young moving groups, which are also PTTS candidates.\\
 
This sample promises to give new insight into various fields of research, such as: {\em i)} the galactic disc heating and the recent local star formation rate, {\em ii)} the discovery of isolated very young stars that are promising targets for planet or debris-disc searches with ground-based or space telescopes, {\em iii)} the evolution of binaries and multiple systems.\\

\begin{acknowledgements}
The authors are grateful to the anonymous referee for useful suggestions. We are grateful to the OHP night assistant staff in conducting our Key Programme, and those of the OAC observatories for their support and help with the observations. This research made use of SIMBAD and VIZIER databases, operated at the CDS, Strasbourg, France. This publication uses ROSAT data. A.~K. also thanks the MEN and ULP for financial support. A partial support from the Italian {\it Ministero dell'Istruzione, Universit\`a e  Ricerca} (MIUR) is also acknowledged.

\end{acknowledgements}

\Online
\begin{appendix}

\section{The data}
\label{Sec:Cat}
The data are published in electronic form only. The tables, available at the CDS, contains all data for the 426 stars we observed at high resolution and discussed in this paper. The master tables (one for S, and SB1; the second one for SB2, and SB3) list all the time-independent data. In particular we have tabulated: the name (RasTyc and SIMBAD; columns \#1 and \#2), Tycho right ascension, and declination (columns \#3 and \#4), $B-V$ colour index (column \#5), ROSAT PSPC count rate, and associated error (columns \#6 and \#7). In addition to these basic data, we also list the parameters derived in this study (but the radial velocity): multiplicity status (column \#8 : S for single or SB1, SB2, and SB3 for single-lined, double-lined, and triple systems, respectively), rotational velocity (column \#9), effective temperature (column \#11), gravity (column \#13), and metallicity (column \#15) with their associated errors (column \#10, \#12, \#14, and \#16, respectively). Finally we also list the measured net H$\alpha$ (column \#19 and \#20) and lithium (column \#21 and \#22) equivalent widths and errors for single and SB1 stars only. For the case where the CCF had many peaks, we list the individual measurements in separate rows and refer to the primary (suffix $a$) as the component with the stronger absorption lines (i.e. the larger area of the CCF dip). The radial velocity measurements and errors are listed in dedicated tables together with the julian date of observation.  

\section{Notes on PMS-like candidates}

We briefly summarize the known characteristics of these stars as recovered from the SIMBAD database and references therein.
All the stars were observed by 2MASS.

\begin{itemize}
\item RasTyc~0221+4600 : classified as a K0 star.
\item RasTyc~0307+3020 : considered as a T Tauri candidate (WTTS) by \cite{1998A&AS..132..173L} on the basis of the ROSAT RASS-BSC \citep{1996IAUC.6420....2V} data selected by high hardness ratios. They quoted a spectral type G5IV. The star was also classified as PMS by \cite{2005A&A...438..769D} in their \textit{PMS stars Proper Motion Catalogue}.
\item RasTyc~0319+4212 : classified as a G5 T Tauri star.
\item RasTyc~0335+3113 : classified as a G0 star. \cite{1998A&AS..132..173L} considered this star as WTTS candidate with spectral type G5IV. Also considered as PMS in the \cite{2005A&A...438..769D} catalogue.
\item RasTyc~0348+5202 : quoted as a double or multiple G6IV star \citep{2002BaltA..11..153B} while classified as K0V in \cite{1984ApJS...56..257J} spectral atlas. \cite{1992ApJS...81..865S} had taken a low resolution spectrum in the visible and near IR included in the {\it New library of stellar optical spectra}. 
\item RasTyc~1825+6450 :  classified as G5IV by \cite{1983ApJS...52..121G} and identified as a new periodic variable from the Hipparcos epoch photometry \citep{2002MNRAS.331...45K}. In their kinematic and dynamical studies, \cite{2005A&A...430..165F} found rich small-scale structure, with several clumps corresponding to various streams and superclusters. This star is part of their sample but it is not clear to what group they decided to attach it.
\item RasTyc~2100+4530 : quoted as an F8 star. Observed at low resolution by \cite{1997A&A...318..111M} in the framework of an identification programme of RASS galactic plane sources.
\end{itemize}

\end{appendix}
\onlfig{11}{
\begin{figure*}[htp]
\includegraphics[width=9.0cm]{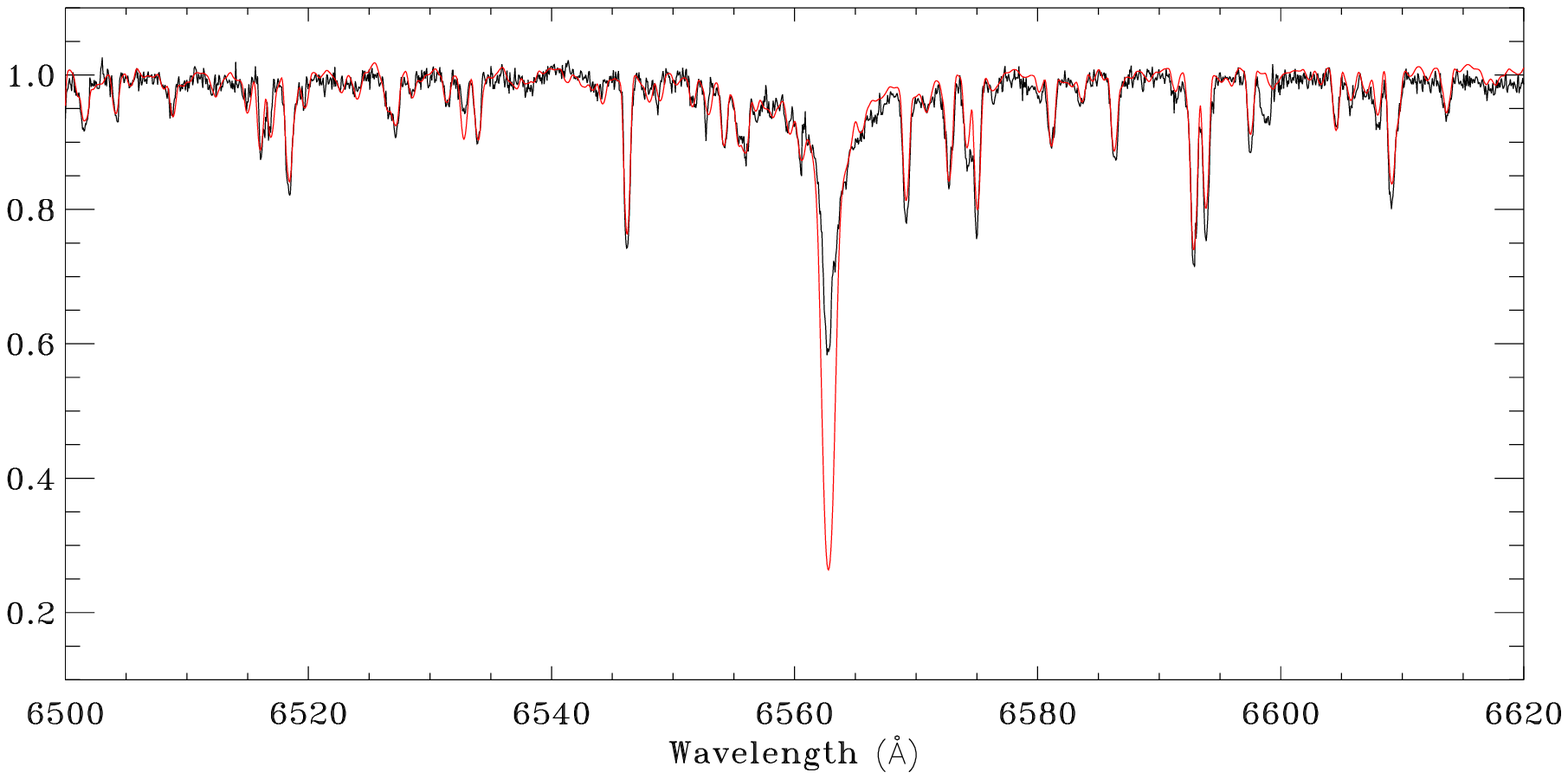}
\includegraphics[width=9.0cm]{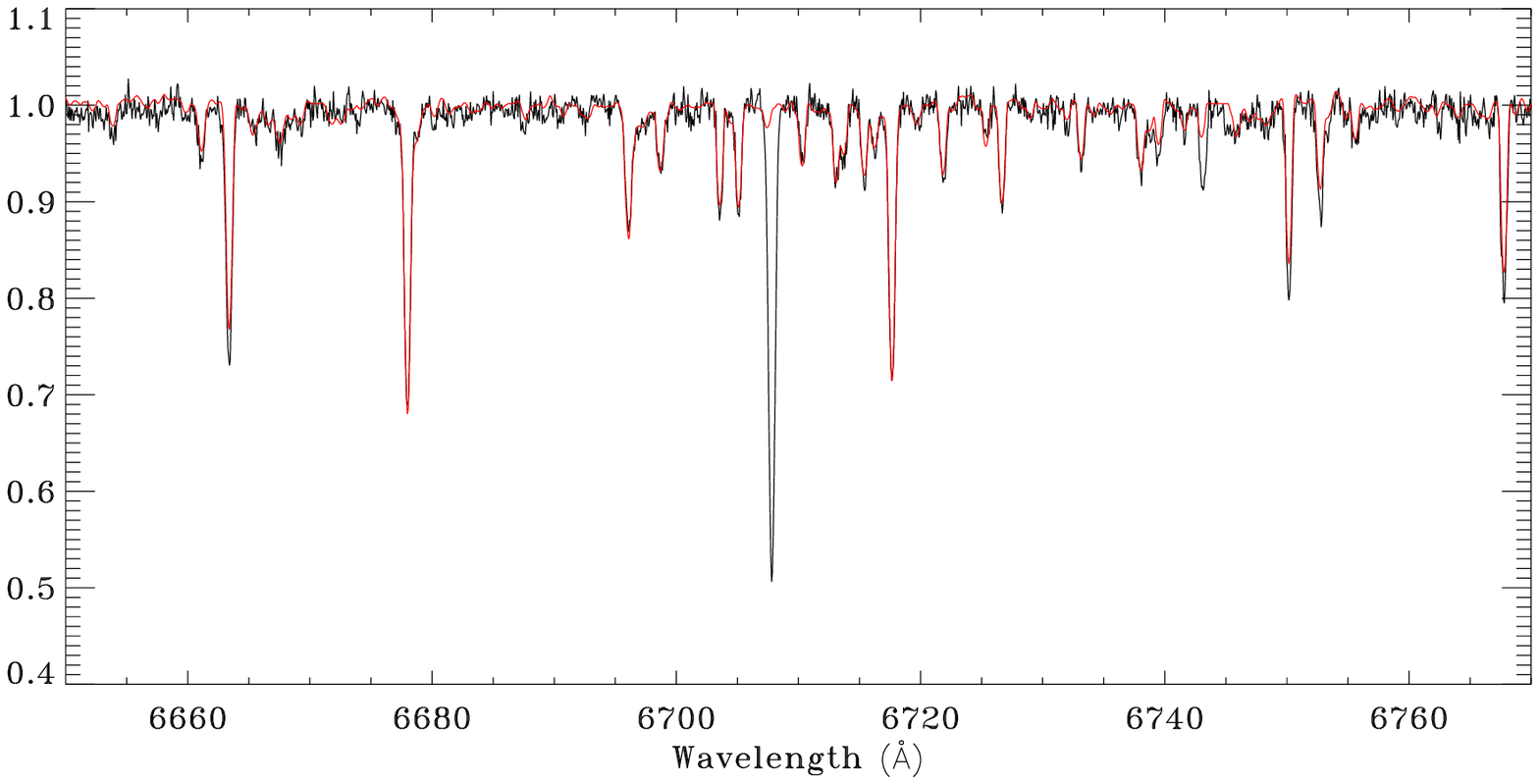}
\begin{picture}(180.,40.)
\end{picture}
\includegraphics[width=9.0cm]{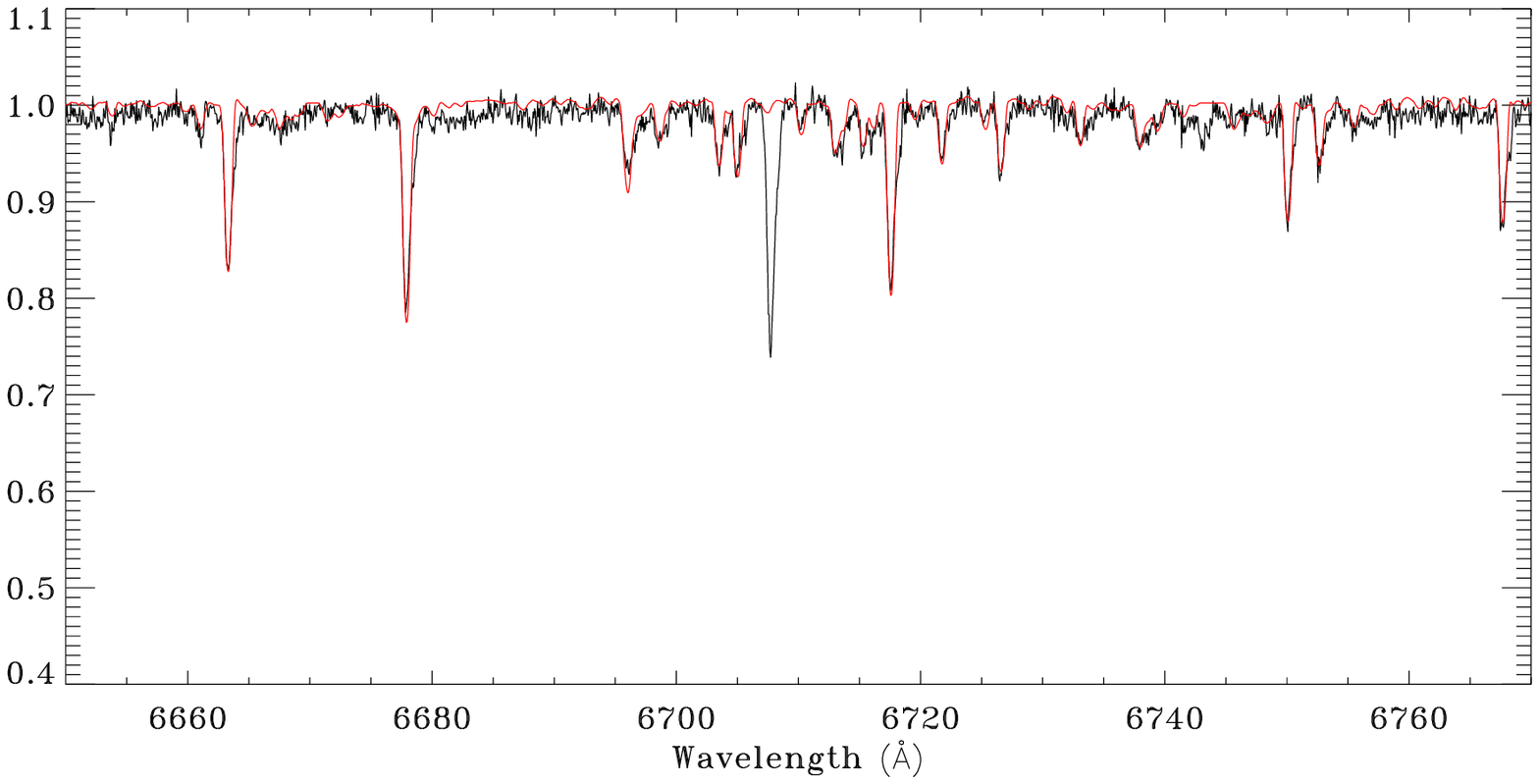}
\includegraphics[width=9.0cm]{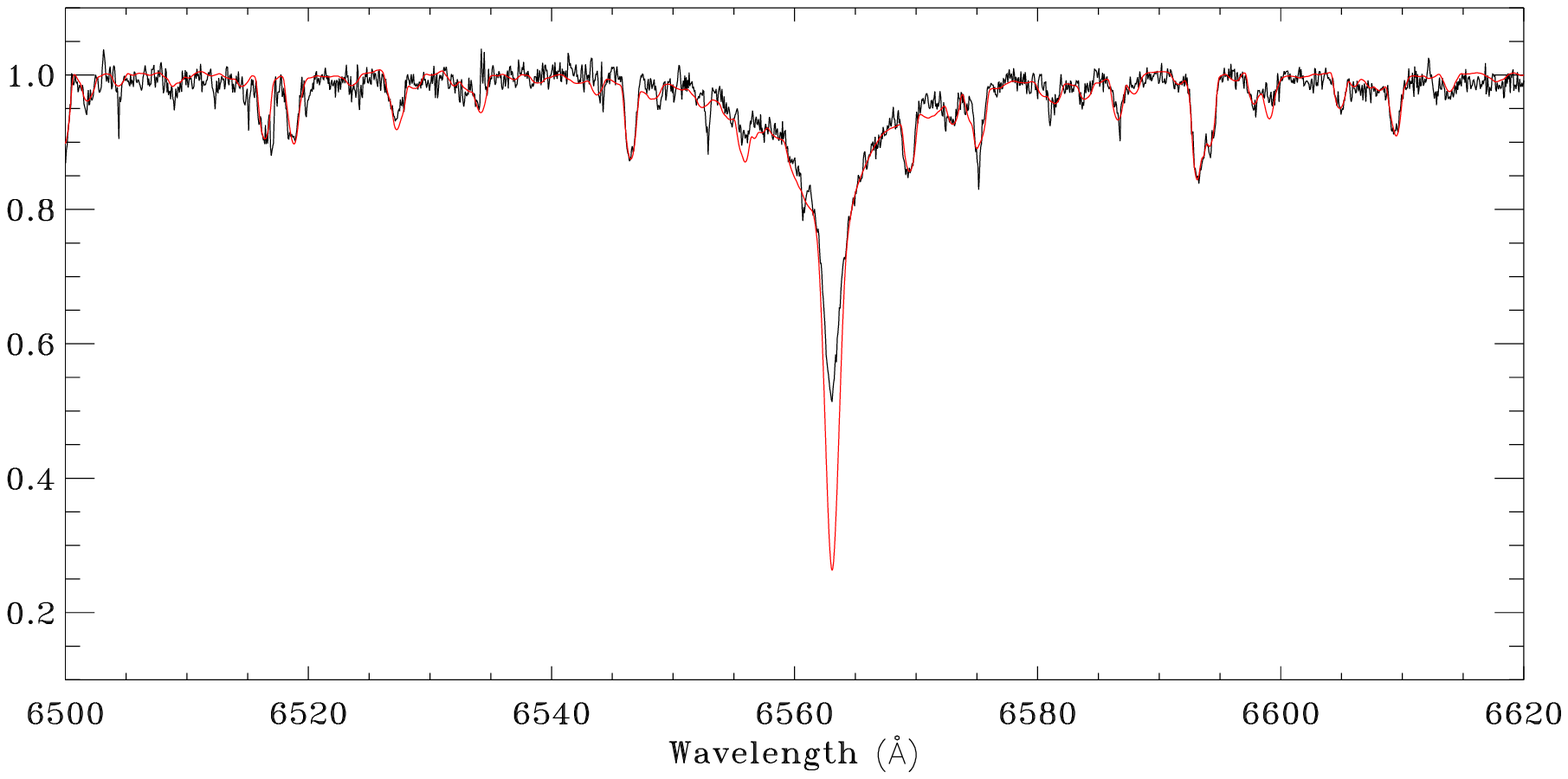}
\includegraphics[width=9.0cm]{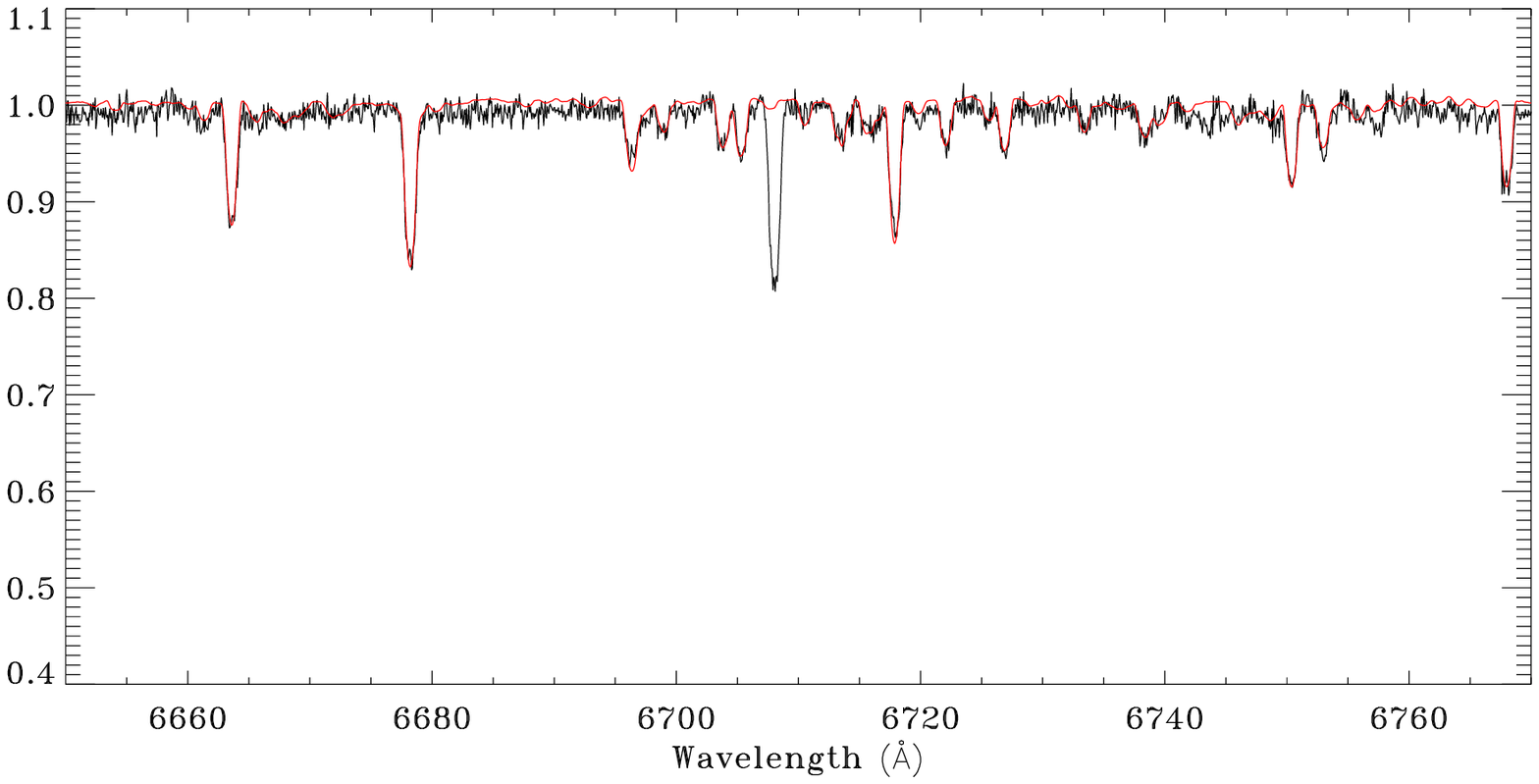}
\includegraphics[width=9.0cm]{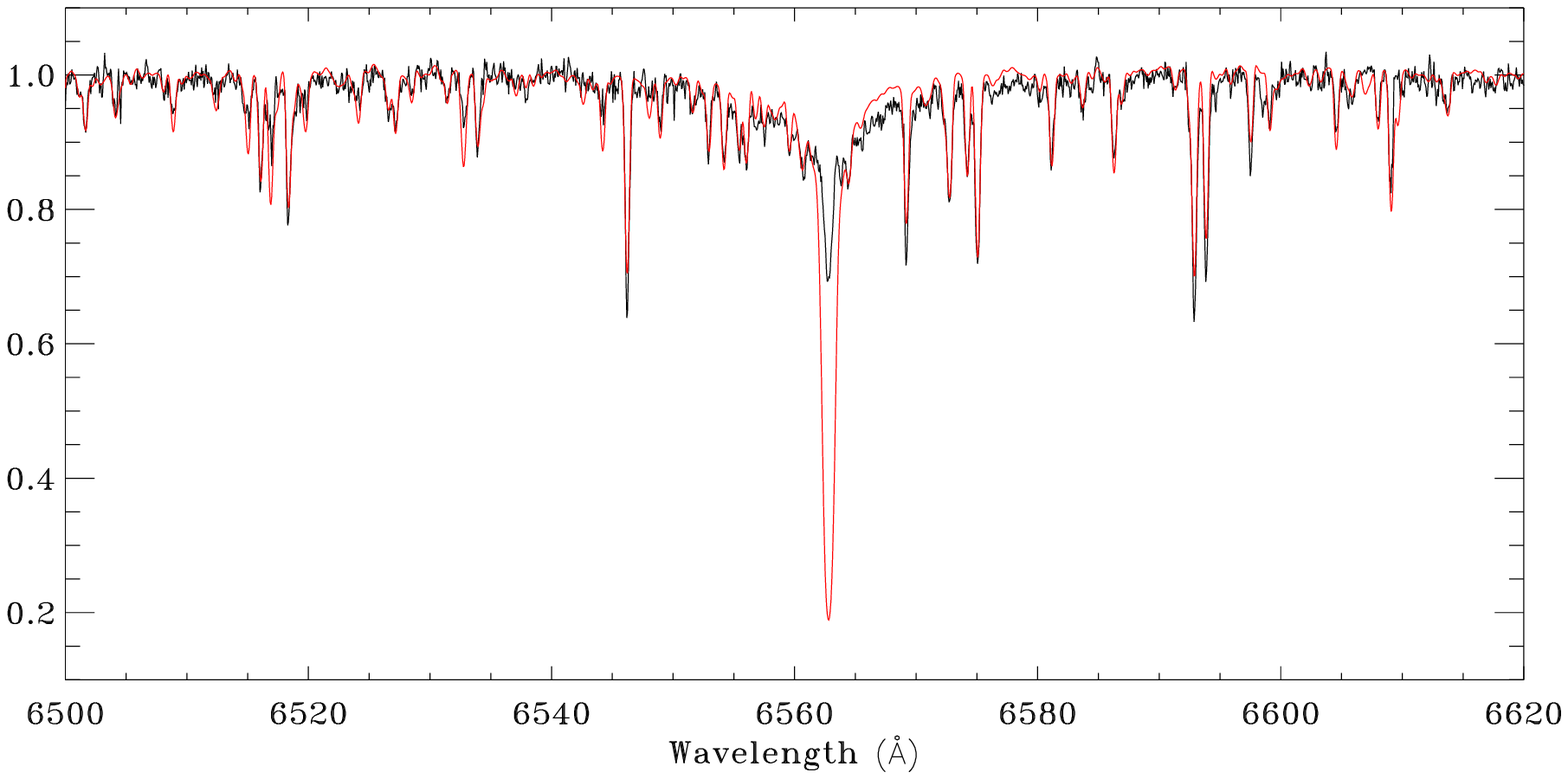}
\includegraphics[width=9.0cm]{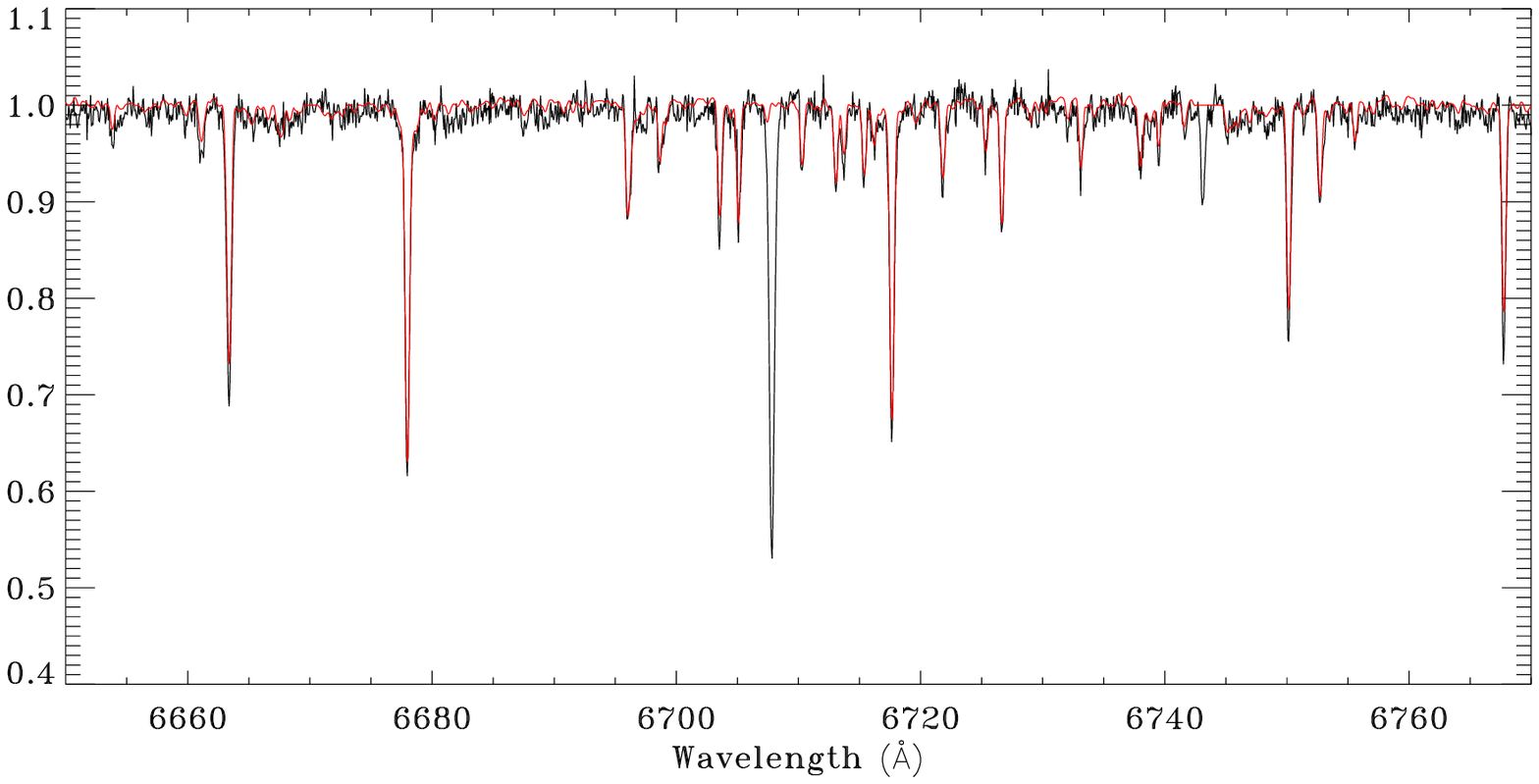}
\includegraphics[width=9.0cm]{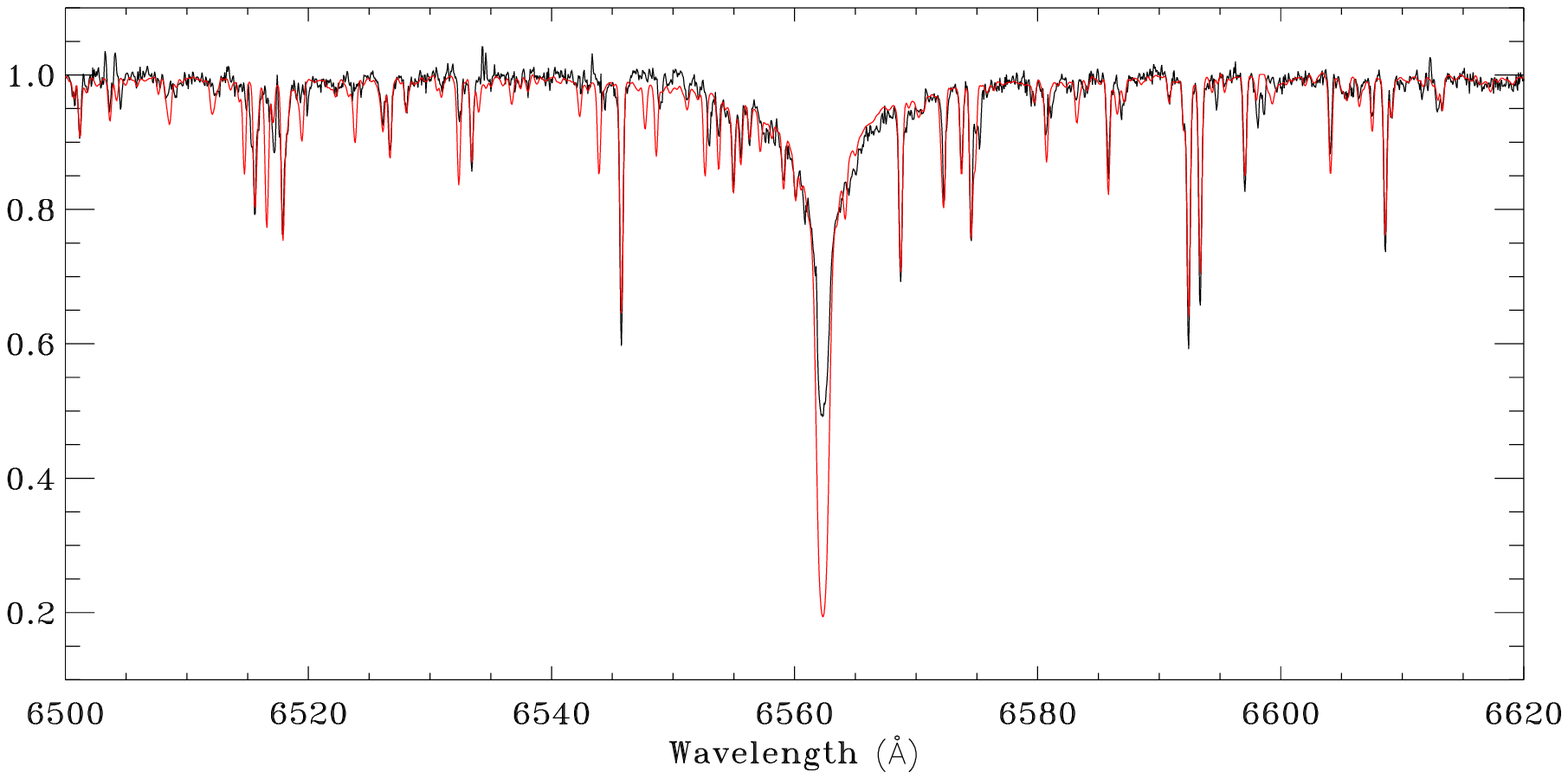}
\includegraphics[width=9.0cm]{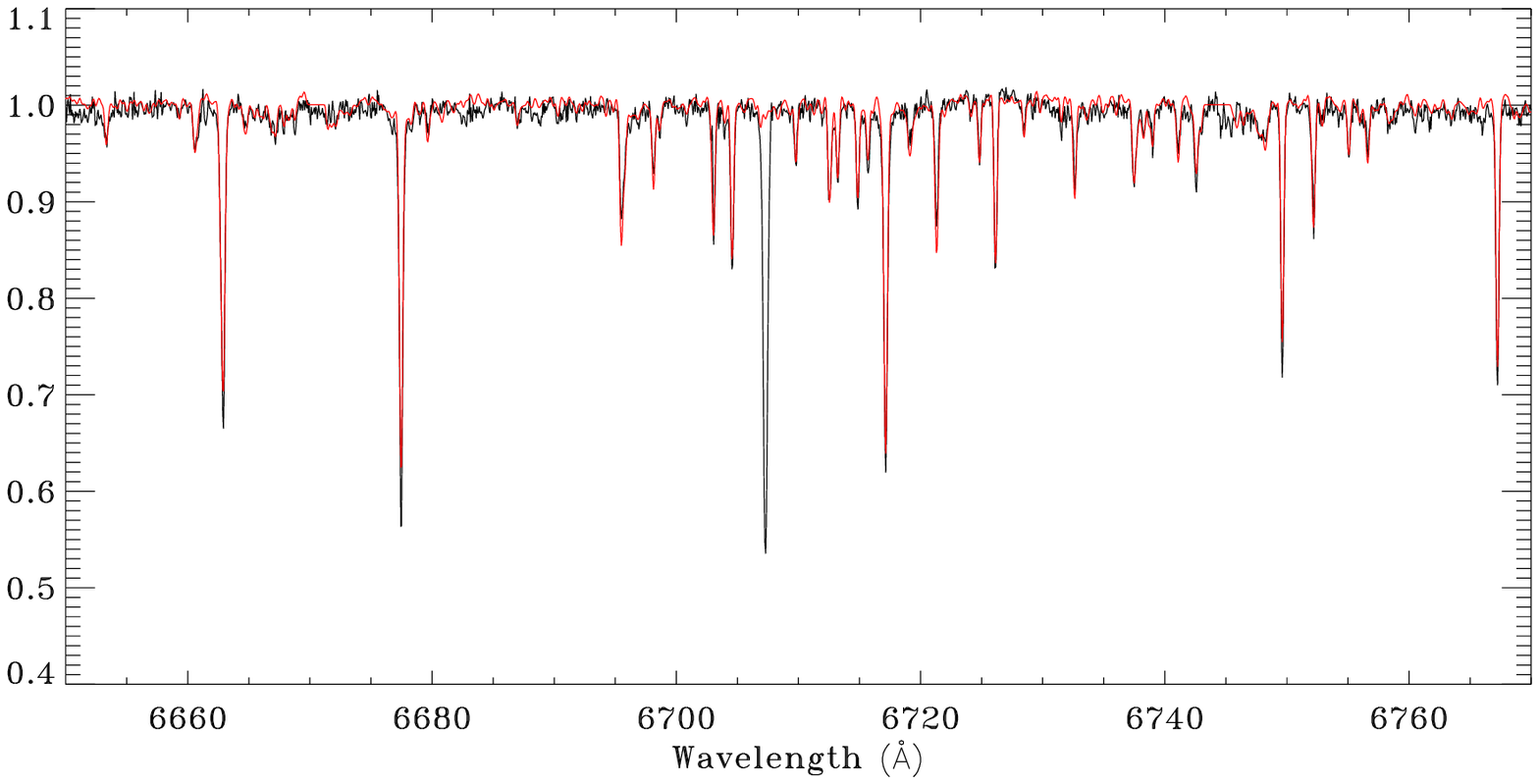}
\caption{H$\alpha$ (\textit{left panel}) and lithium (\textit{right panel}) spectra (black solid line) of the new moving-groups PTTS candidates RasTyc~0221+4600, RasTyc~0307+3020, RasTyc~0335+3113, RasTyc~0348+5202, and RasTyc~2100+4530 (from top to bottom), together with the synthetic spectrum built up with the reference spectra broadened at the $v\sin i$ of the star's components and Doppler-shifted according to its RV (thin red line).}
\label{Fig:PTTSspec}
\end{figure*}
}
\onlfig{12}{
\begin{figure*}[ht]
\includegraphics[width=9.0cm]{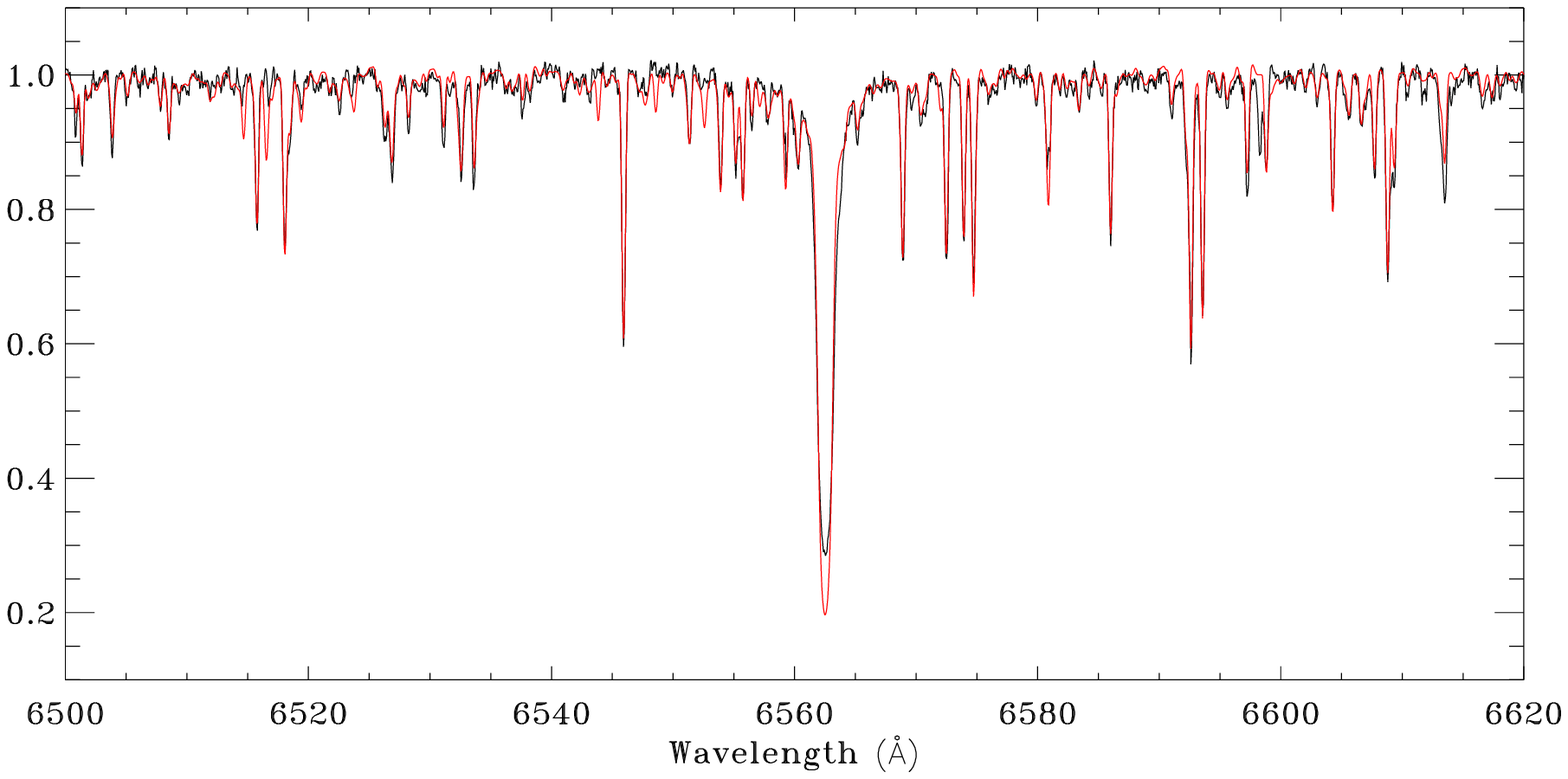}
\includegraphics[width=9.0cm]{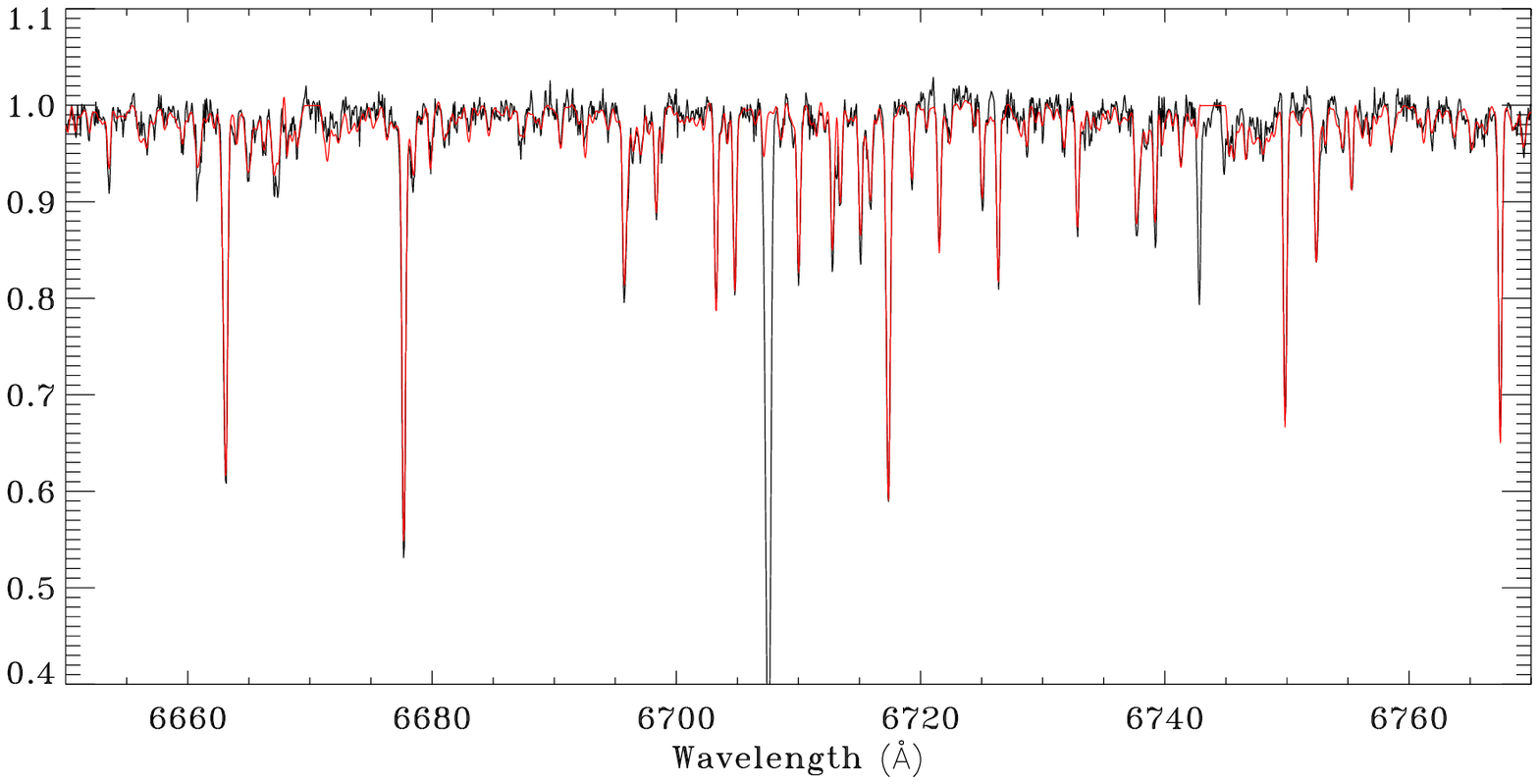}
\includegraphics[width=9.0cm]{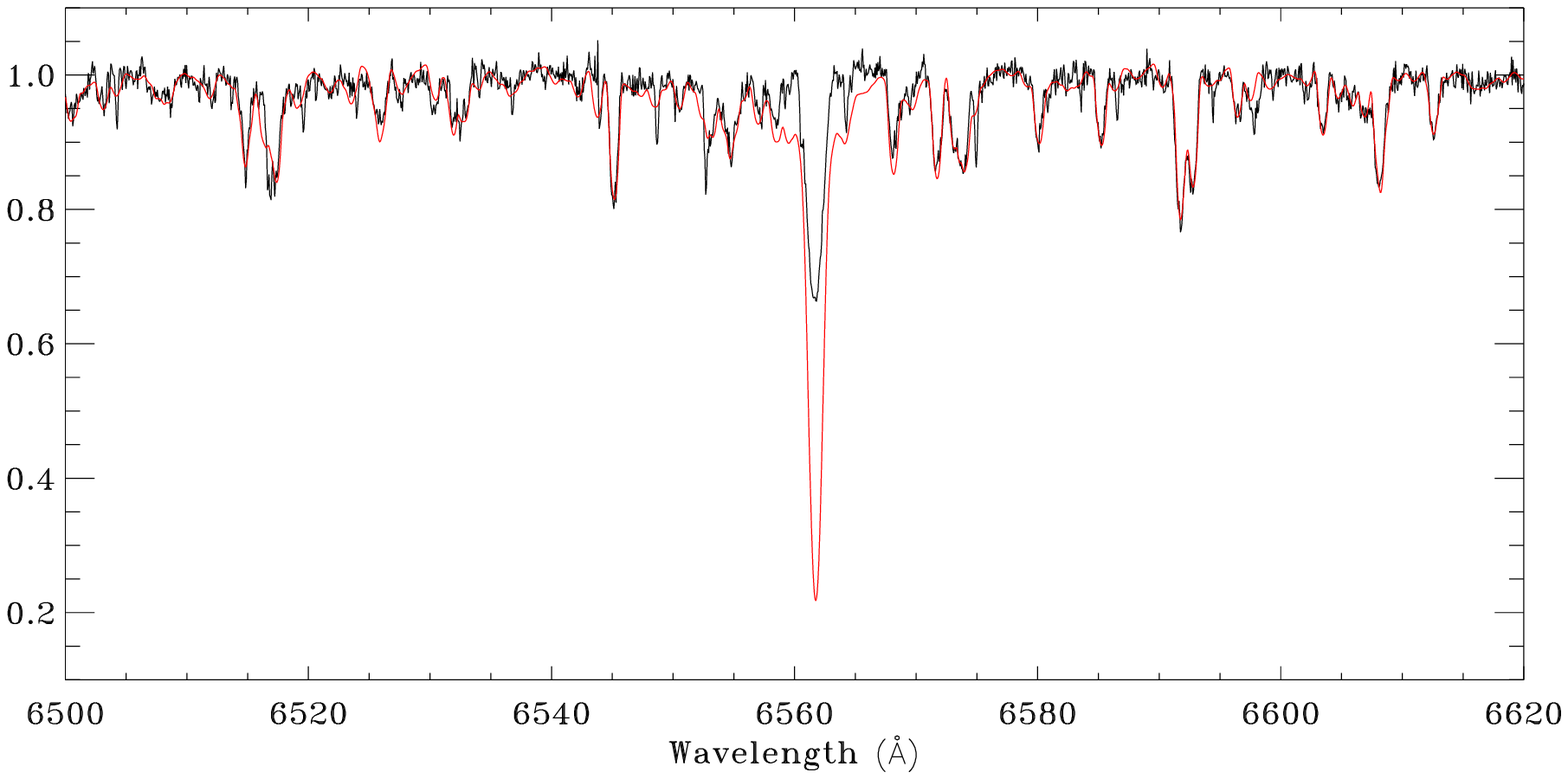}
\includegraphics[width=9.0cm]{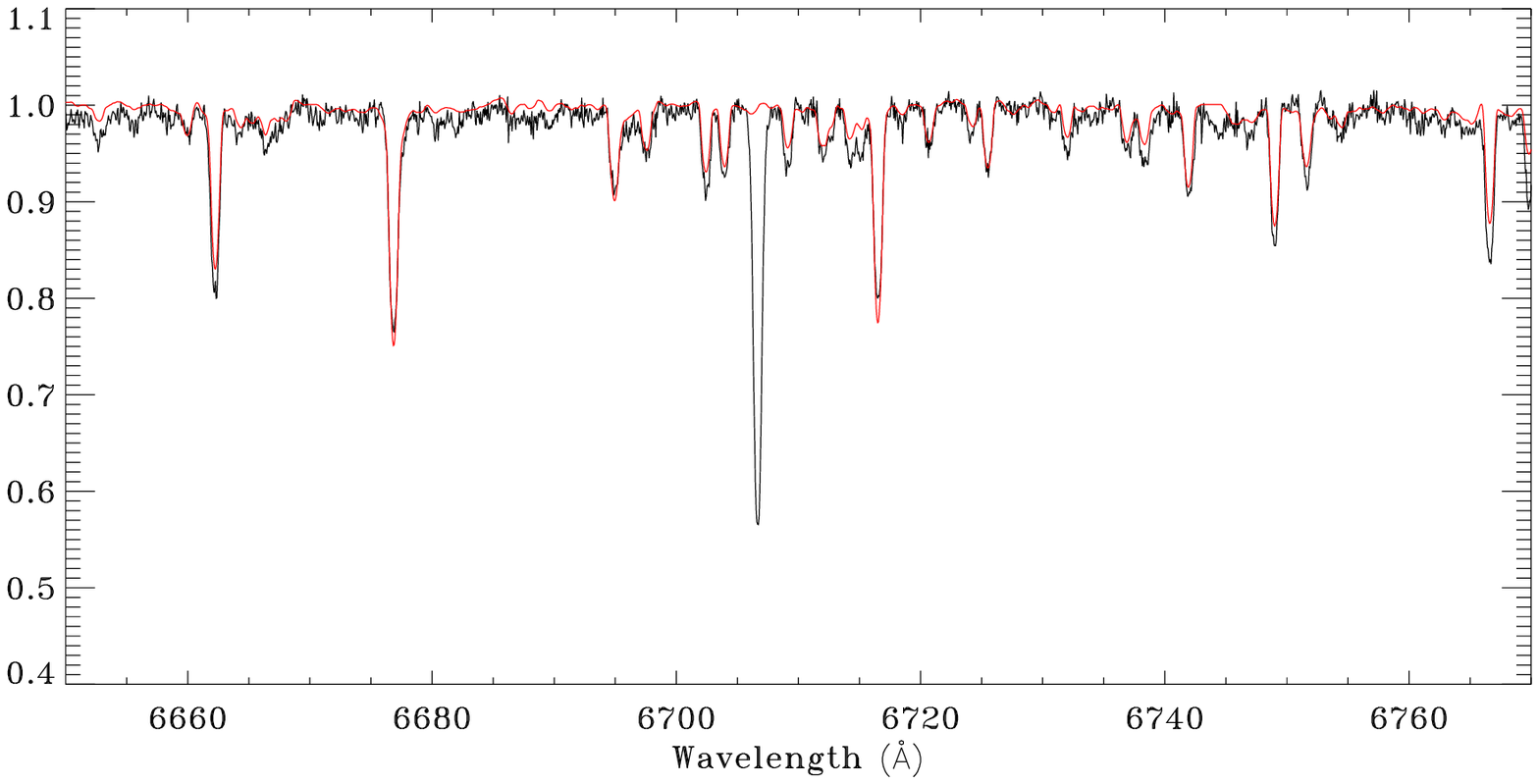}
\caption{H$\alpha$ (\textit{left panel}) and lithium (\textit{right panel}) spectra (black solid line) of lithium-rich giants RasTyc~0319+4212 and RasTyc~1825+6450 (from top to bottom), together with the synthetic spectrum built up with the reference spectra broadened at the $v\sin i$ of the star's components and Doppler-shifted according to its RV (thin red line).}
\label{Fig:LiRichGiantspec}
\end{figure*}
}

\end{document}